\pdfoutput=1
\documentclass[english,onecolumn, draftclsnofoot]{IEEEtran}
\usepackage[T1]{fontenc}
\usepackage[latin9]{inputenc}
\usepackage{babel}
\usepackage{amsmath}
\usepackage{amsthm}
\usepackage{amssymb}
\usepackage{stmaryrd}
\usepackage[pdftex]{graphicx}
\usepackage[pdftex, unicode=true,
 bookmarks=true,bookmarksnumbered=true,bookmarksopen=true,bookmarksopenlevel=1,
 breaklinks=false,pdfborder={0 0 1},backref=false,colorlinks=false]
 {hyperref}
\hypersetup{pdftitle={Exact Dimensionality Reduction for Partial Line Spectra Estimation Problems},
 pdfauthor={Maxime Ferreira Da Costa and Wei Dai},
 pdfsubject={Exact Dimensionality Reduction for Partial Line Spectra Estimation Problems},
 pdfkeywords={Sampling theory; line spectral estimation; super-resolution; sub-Nyquist sampling; multirate sampling; convex optimization; dimensionality reduction}}

\makeatletter

\newcommand{\noun}[1]{\textsc{#1}}

\numberwithin{equation}{section}
\numberwithin{figure}{section}
\numberwithin{table}{section}
\theoremstyle{plain}
\newtheorem{thm}{\protect\theoremname}[section]
\theoremstyle{plain}
\newtheorem{lem}[thm]{\protect\lemmaname}
\theoremstyle{plain}
\newtheorem{prop}[thm]{\protect\propositionname}
\theoremstyle{definition}
\newtheorem{defn}[thm]{\protect\definitionname}
\theoremstyle{remark}
\newtheorem{rem}[thm]{\protect\remarkname}
\theoremstyle{plain}
\newtheorem{cor}[thm]{\protect\corollaryname}

\makeatother

\providecommand{\corollaryname}{Corollary}
\providecommand{\definitionname}{Definition}
\providecommand{\lemmaname}{Lemma}
\providecommand{\propositionname}{Proposition}
\providecommand{\remarkname}{Remark}
\providecommand{\theoremname}{Theorem}

\begin{document}

\title{Exact Dimensionality Reduction for\\
Partial Line Spectra Estimation Problems}

\author{\IEEEauthorblockN{Maxime Ferreira Da Costa and Wei Dai}\linebreak{}
\IEEEauthorblockA{Department of Electrical and Electronic Engineering, Imperial College
London, United Kingdom\linebreak{}
Email: \{maxime.ferreira, wei.dai1\}@imperial.ac.uk}}
\maketitle
\begin{abstract}
Line spectral estimation theory aims to estimate the off-the-grid
spectral components of a time signal with optimal precision. Recent
results have shown that it is possible to recover signals having sparse
line spectra from few temporal observations via the use of convex
programming. However, the computational cost of such approaches remains
the major flaw to their application to practical systems. This work
investigates the recovery of spectrally sparse signal from low-dimensional
partial measurements. It is shown in the first part of this paper
that, under a light assumption on the sub-sampling matrix, the partial
line spectral estimation problems can be relaxed into a low-dimensional
semidefinite program. The proof technique relies on a novel extension
of the Gram parametrization to subspaces of trigonometric polynomials. 

The second part of this work focuses on the analysis of two particular
sub-sampling patterns: multirate sampling and random selection sampling.
It is shown that those sampling patterns guarantee perfect recovery
of the line spectra, and that the reconstruction can be achieved in
a poly-logarithmic time with respect to the full observation case.
Moreover, the sub-Nyquist recovery capabilities of such sampling patterns
are highlighted. The atomic soft thresholding method is adapted in
the presented framework to estimate sparse spectra in noisy environments,
and a scalable algorithm for its resolution is proposed. 
\end{abstract}


\global\long\def\trans{\mathsf{T}}

\global\long\def\herm{*}

\global\long\def\splus{\mathrm{{\scriptscriptstyle +}}}

\global\long\def\linmeasure{\mathcal{L}}

\global\long\def\range{{\rm range}}

\global\long\def\sign{{\rm sign}}

\section{Introduction}

\IEEEPARstart{C}{ompressed} sensing techniques have proven to be
of great interests for detecting, estimating and denoising sparse
signals lying on discrete spaces. On the practical side, the applications
of sparse modeling are many: single molecule imaging via fluorescence,
blind source separation in speech processing, precise separation of
multiple celestial bodies in astronomy, or super-resolution radaring,
are among those. However, the discrete gridding required by the compressed
sensing framework weaken the recovery performances, and more precisely
the system \emph{resolution}: the required minimal separation between
two components of the sparse signal to be efficiency distinguished
by an observation process.

In the recent years, a particular enthusiasm has been placed on solving
sparse linear inverse problems over continuous dictionaries. This
aims to recover the smallest finite subset of components generating
a signal, and lying in a continuous space, by discrete observations
of this signal distorted by a kernel function. Considering such approach
raises new theoretical and practical concerns, in particular, those
problems are commonly infinitely ill-posed.

This paper will discuss the spectral spikes recovery problem, also
known as line spectrum estimation problem, which is probably one of
the most fundamental and important illustration of sparse modeling
over continuous spaces. For the spectral case, a complex time signal
$x$ is said to follow the $s$-spikes model if and only if it reads
\begin{equation}
\forall t\in\mathbb{R},\quad x\left(t\right)=\sum_{r=1}^{s}\alpha_{r}e^{i2\pi\xi_{r}t},\label{eq:SpikesModel}
\end{equation}
whereby $\Xi=\left\{ \xi_{r},\;r\in\left\llbracket 1,s\right\rrbracket \right\} $
is the ordered set containing the $s$ spectral components generating
the signal $x$, and $\alpha=\left\{ \alpha_{r},\;r\in\left\llbracket 1,s\right\rrbracket \right\} $
the one of their associated complex amplitudes. 

In the \emph{total observation} framework, i.e. when observing $n\in\mathbb{N}$
uniform samples of the form $y\left[k\right]=x\left(\frac{k}{f}\right)$
for some sampling frequency $f\in\mathbb{R}^{+}$, the frequency estimation
problem is naturally defined as building a consistent estimator $\left(\bar{\Xi},\bar{\alpha}\right)$
of the parameters $\left(\Xi,\alpha\right)$, that are supposed to
be unknown, of the time signal $x$ based on the knowledge of $y\in\mathbb{C}^{n}$.
This problem is obviously ill-posed, and since no assumption is a
priori made on the number of frequencies $s$ to estimate, there are
infinitely many pairs $\left(\bar{\Xi},\bar{\alpha}\right)$ that
are consistent with the observations. As for illustration purpose,
the discrete Fourier transform of the observation vector $y$ forms
a consistent spectral representation of the signal $x$ by $n$ spectral
spikes at locations $\bar{\xi}_{k}=\frac{k}{n}f$ in the frequency
domain. However, this representation has generally no reason to be
sparse, in the sense that a time signal $x$ drawn from the $s$-spikes
model will be represented by $n>s$ non-null spectral coefficients;
unless the all the elements of $\Xi$ exactly belongs to the spectral
grid $\left\{ \frac{k}{n}f,\,k\in\mathbb{\mathbb{Z}}\right\} $.

Among all those consistent estimators, the one considered to be optimal,
in the sparse recovery context, is the one returning the sparsest
spectral distribution, i.e., the one that outputs a spectral support
$\bar{\Xi}_{0}$ achieving the smallest cardinality $\bar{s}_{0}$.
Consequently, under total observation, by denoting $\hat{x}$ the
spectrum of $x$, the optimal spectral estimator $\hat{x}_{0}$ of
$\hat{x}$ can be written as the output of an optimization program
of the form
\begin{align}
\hat{x}_{0} & =\arg\min_{\hat{x}\in D_{1}}\left\Vert \hat{x}\right\Vert _{0}\label{eq:GenericL0Problem}\\
\text{subject to} & \phantom{\;=\;}y=\mathcal{F}_{n,f}\left(\hat{x}\right),\nonumber 
\end{align}
where $\left\Vert \cdot\right\Vert _{0}$ represents the limit of
the $p$ pseudo-norm towards $0$, counting the cardinality of the
support. $D_{1}$ denotes the space of absolutely integrable spectral
distributions, and $\mathcal{F}_{n,f}$ is the inverse discrete time
Fourier transform for the sampling frequency $f\in\mathbb{R}^{+}$
defined by,
\begin{align}
\mathcal{F}_{n,f}:\;D_{1} & \to\mathbb{C}^{n}\label{eq:Def-InverseDTFT}\\
\hat{x} & \mapsto q:\,q\left[k\right]=\int_{\mathbb{R}}e^{i2\pi\frac{\xi}{f}k}\mathrm{d}\hat{x}\left(\xi\right),\;\forall k\in\left\llbracket 0,n-1\right\rrbracket .\nonumber 
\end{align}
In case of absence of ambiguity on $f$, its notation will be simplified
to $\mathcal{F}_{n}$.

Program (\ref{eq:GenericL0Problem}) is non-convex, and the combinatorial
nature of ``$L_{0}$'' minimization leaves the direct formulation
of this problem practically unsolvable. A commonly proposed workaround
consists in analyzing the output of a relaxed problem, obtained by
swapping the cardinality cost function $\left\Vert \cdot\right\Vert _{0}$
into a minimization of the total-variation norm over the spectral
distribution domain $\left\Vert \cdot\right\Vert _{\textrm{TV}}$
, defined by
\[
\left\Vert \hat{x}\right\Vert _{\textrm{TV}}=\sup_{f\in C\left(\mathbb{R}\right),\left\Vert f\right\Vert _{\infty}\leq1}\Re\left[\int_{\mathbb{R}}\overline{f\left(\xi\right)}\mathrm{d}\hat{x}\left(\xi\right)\right],
\]
where $C\left(\mathbb{R}\right)$ denotes the space of continuous
complex functions of the real variable. The total-variation norm can
be interpreted as an extension of the $L_{1}$ norm to the distribution
domain. This relaxation leads to the formulation of the convex program
\begin{align}
\hat{x}_{\textrm{TV}} & =\arg\min_{\hat{x}\in D_{1}}\left\Vert \hat{x}\right\Vert _{\textrm{TV}}\label{eq:GenericL1Problem}\\
\text{subject to} & \phantom{\;=\;}y=\mathcal{\mathcal{F}}_{n,f}\left(\hat{x}\right).\nonumber 
\end{align}

Sufficient conditions for the tightness of this relaxation have been
successfully addressed in \cite{Candes2014a,Candes2012,Tang2013,Fernandez-granda2015}:
Problem (\ref{eq:GenericL1Problem}) is known to output a spectral
distribution $\hat{x}_{\textrm{TV}}$ that is equal to the optimal
solution $\hat{x}_{0}$ of the original Problem (\ref{eq:GenericL0Problem})
under the mild separation assumption between the spikes in the frequency
domain

\begin{equation}
\Delta_{\mathbb{T}}\left(\frac{1}{f}\Xi\right)\geq\frac{2.52}{n-1},\label{eq:MinimalSeparabilityCond}
\end{equation}
provided that the number of measurements $n$ is greater than some
constant, and whereby $\Delta_{\mathbb{T}}\left(\cdot\right)$ is
the set minimal warp around distance over the elementary torus $\mathbb{T}=\left[0,1\right)$
defined by
\[
\forall\Omega\subset\mathbb{R},\quad\Delta_{\mathbb{T}}\left(\Omega\right)=\min\left\{ {\rm frac}\left(\nu-\nu^{\prime}\right),\;\left(\nu,\nu'\right)\in\Omega^{2},\nu\neq\nu^{\prime}\right\} ,
\]
and whereby $\mathrm{frac}\left(\cdot\right)$ denotes the fractional
part of any real number. Nevertheless, the estimate $\hat{x}_{\textrm{TV}}=\hat{x}_{0}$
will correspond to true distribution $\hat{x}$ only if the Nyquist
criteria is met, since an ambiguity modulo $f$ stands in the spectral
domain due to the aliasing effect generated by the uniform sampling
process.

\subsection*{Related work on line spectral estimation}

Up to recent years, most of the approaches to recover the off-the-grid
spikes generating sparse signals where based on \emph{subspace construction}
methods. It is the case of the popular and proven \noun{Music \cite{Stoica1989}}
and \noun{Esprit \cite{Roy1989}} methods, building tap delayed subspaces
from the measurements, and making use of their low rank properties
to locate the frequencies while denoising signals. A more recent method
\cite{Blu2008}, based on annihilating and Cadzow filtering, describes
an algebraic framework to estimate the set of continuous frequencies.
If many of those methods have been shown to build consistent estimates,
little is known about the theoretical spectral accuracy of those estimates
under noisy observations.

The interest for approaching the line spectrum estimation problem
under the lens of \emph{convex optimization} has been increasing after
that the recent work \cite{Candes2014a} established the optimality
of convex relaxation under the previously discussed conditions. It
has been shown in \cite{Tang2013} that such optimality still holds
with high probability when extracting at random a small number of
observations and discarding the rest of it.

The convex approach has been proven to be robust to noise, achieving
near optimal mean-square error in Gaussian noise \cite{Tang2013a}
under full measurements. The dispersion in $L_{1}$ norm in the time
domain has been bounded in \cite{Candes2012} for an arbitrary noise
distribution. The sufficient separability  criterion on the spikes
has been enhanced in \cite{Fernandez-granda2015}, and authors of
\cite{Bhaskar2013} demonstrated that the estimated time signal converges
in quadratic norm to the time signal $x$ without any spectral separability
conditions when the number of observations grows large.

The line spectral estimation problem is a practically important sub-case
for the wider theory for pulse stream deconvolution. A general analysis
of this framework is presented in \cite{Duval2015}, sufficient conditions
of the tightness of the convex relaxation approach have been proposed
\cite{Bendory2016511}, while \cite{Tang2015} provides necessary
ones. Authors of \cite{Schiebinger2015} proved that the deconvolution
of spikes is possible without separation assumption for a broad class
of distortion kernels, including the Gaussian one. On the computational
side, several algorithms have been proposed to bridge the high computational
cost of solving the relaxed Program (\ref{eq:GenericL1Problem}),
including a space discretization approach in \cite{Tang}, and an
enhanced gradient search for sparse inverse problems in \cite{Boyd2015}.

Many extensions of the spectral spikes model have been studied. The
recent works \cite{Chi2015,Yang2015} extend to the case of multi-dimensional
spikes, proving the efficiency of convex relaxations, although the
resolution degrades with the order of the model. Estimation from multiple
measurement vectors (MMV) has been proposed in \cite{Li2014a,Yang2014}.
More generic models involving spectral deconvolution of spikes from
unknown kernels have been studied in \cite{Yang2016}. 

Other relaxation approaches to recover the spectral spikes exist in
the literature. In \cite{Cai2016}, a nuclear norm minimization over
the set of Hankel matrices were proved to return exact estimates without
the need of any separation condition. Authors of \cite{7478129} recently
considered a relaxation using log-penalty functions achieving better
empirical performances. However the robustness of those estimators
to noisy environments remains unexplored. 

Finally, on the practical side, the super-resolution theory of spikes
has found application to super-resolution fluorescence microscopy
and more recently to super-resolution radar imaging \cite{Heckel2014}.

We emphasize on the fact that the cited studies address the line spectrum
estimation problem under full observations $y\in\mathbb{C}^{n}$ acquired
uniformly for some sampling frequency.

\subsection*{Focus and organization of this paper}

If line spectral search is a theoretically promising approach to recover
sparse spectra with very high precision, the computational complexity
of the convex relaxation approach (\ref{eq:GenericPartialL1}) remains
the principal flaw to its use in practice. A direct approach to recover
the spectra $\hat{x}_{\textrm{TV}}$ using classic convex solvers
grows as $\mathcal{O}\left(n^{7}\right)$ in the number of measurements
$n$ and becomes unrealistic when dealing with more than a few hundred
of them.

This work aims to address the complexity issue by recovering the spectrum
of the probed time signal $x$ via \emph{partial observations $y\in\mathbb{C}^{m}$},
obtained as linear combinations of the output of a uniform sampler
$y_{\textrm{raw}}\in\mathbb{C}^{n}$, such that $y=My_{\textrm{raw}}$.
The sub-sampling matrix $M\in\mathbb{C}^{m\times n}$ defines the
linear combinations to apply on the raw output of the uniform sampler.
We show that, under an unrestrictive admissibility condition on the
sub-sampling matrix $M$, the line spectral estimation problem can
be reformulated as a semidefinite program of dimension $m+1$. Moreover
we study some categories of sub-sampling matrix and derive sufficient
conditions for optimal recovery of the spectrum $\hat{x}$ of the
probed signal from sub-Nyquist sampling rates. We show that our approach
can bring \emph{orders of magnitude} changes to the computational
complexity of the recovery, turning the standard polynomial time algorithm
into equivalent ones of \emph{poly-logarithmic} orders.

The present work is essentially organized in three parts. In the first
part, Section \ref{sec:Exact-Dimensionality-Reduction} presents the
\emph{partial line spectral estimation} framework and states generic
conditions for the recoverability of any time signal $x$ following
Model (\ref{eq:SpikesModel}). It further introduces our main result
in Theorem \ref{thm:Dual-CompactSDP-Equivalence}, establishing the
recoverability of $x$ from the output of a semidefinite program of
dimension $m+1$. An explicit formulation of this program is provided
for the remarkable case of so called selection matrices.

The second part of this work studies the recoverability of $x$ from
partial measurement acquired through a sub-sampling matrix $M$ having
a selection based structure. Two selection patterns are studied in
details. The first one is presented in Section \ref{sec:Multirate-sampling-systems},
and treat the case where the output $y\in\mathbb{C}^{m}$ is generated
by a multirate sampling systems: a system formed by a set of uniform
samplers working at potentially different delays and frequencies.
It is shown in Theorem \ref{thm:MRSSDualCertifiability} that, under
a common alignment property, involving certain conditions on the rates
and the delays between the samplers, the output of relaxed approach
to the line spectral estimation is tight. Furthermore, the sub-Nyquist
recovery capabilities of the studied framework are highlighted, and
the complexity gain of using such sampling model is discussed. Section
\ref{sec:Random-sub-sampling} presents the random selection sub-sampling
model firstly introduced in \cite{Tang2013} and shows that it can
be used to reconstruct signal following the spikes model in a poly-logarithmic
computational time.

In the last part of this paper, we address in Section \ref{sec:SpectralEstimationInNoise}
the estimation problem from noisy measurements by extending the atomic
soft thresholding (AST) method proposed in \cite{Tang2013a} to our
observation framework. A fast and scalable algorithm based on the
alternative direction method of multipliers (ADMM) is presented in
Section \ref{sec:ADMM} to estimate the spectral spikes from partial
sampling. Finally, Section \ref{sec:Proof-of-Dimension-Reduction}
presents a detailed proof of Theorem \ref{thm:Dual-CompactSDP-Equivalence}
that relies on an elegant extension of the Gram parametrization property
of trigonometric polynomials to subspaces of polynomials.

\section{Dimensionality reduction for partially observed systems\label{sec:Exact-Dimensionality-Reduction}}

\subsection{Problem setup}

We consider the estimation problem of a continuous time signal $x$
following the spikes model (\ref{eq:SpikesModel}) from $m$ partial
observations constructed linearly from the $n$ ($n\geq m$) outputs
of a uniform sampler. This sampler acquires the signal $x$ uniformly
at a given frequency $f\in\mathbb{R}^{+}$. The output of $y_{\textrm{raw}}\in\mathbb{C}^{n}$
of the sampler, before reduction, reads $y_{\textrm{raw}}\left[k\right]=x\left(\frac{k}{f}\right)$
for every sampling index $k\in\left\llbracket 0,n-1\right\rrbracket $.
The observation vector $y\in\mathbb{C}^{m}$ is linked to the uniform
acquisition $y_{\textrm{raw}}$ by the linear relation $y=My_{\textrm{raw}}$
where $M\in\mathbb{C}^{m\times n}$ is the \emph{sub-sampling matrix}
of the system, which is assumed to be known.

As explained before, the line spectrum recovery problem consists in
finding the continuous time signal $x_{0}$ that matches the observations
$y$ while having the sparsest spectral distribution $\hat{x}_{0}$.
In other terms, $\hat{x}_{0}$ has to be composed by the combination
of spikes in the spectral domain of minimal cardinality $s_{0}$.
This ``$L_{0}$'' minimization problem is called \emph{partial line
spectral estimation problem}, and can be described on an analogue
manner to Program (\ref{eq:GenericL0Problem})

\begin{align}
\hat{x}_{M,0} & =\arg\min_{\hat{x}\in D_{1}}\left\Vert \hat{x}\right\Vert _{0}\label{eq:GenericPartialL0}\\
\text{subject to} & \phantom{\;=\;}y=M\mathcal{F}_{n}\left(\hat{x}\right).\nonumber 
\end{align}
The program is known to be NP-hard in the general case due to the
combinatorial search imposed by the ``$L_{0}$'' minimization. Therefore,
we naturally introduce the total-variation counterpart to this problem
in the same manner than (\ref{eq:GenericL1Problem}), leading to

\begin{align}
\hat{x}_{M,\textrm{TV}} & =\arg\min_{\hat{x}\in D_{1}}\left\Vert \hat{x}\right\Vert _{\textrm{TV}}\label{eq:GenericPartialL1}\\
\text{subject to} & \phantom{\;=\;}y=M\mathcal{F}_{n}\left(\hat{x}\right).\nonumber 
\end{align}

In the presented work, we address two fundamental issues arising from
the formulation of the convex formulation (\ref{eq:GenericPartialL1}):
\begin{itemize}
\item \emph{Computational complexity}: Can one solve this convex problem
in a computational time depending only on the dimension of the observations
$m$?
\item \emph{Recoverability}: Can one find sub-sampling matrices $M$ guarantying
the recoverability of the signal $x$ (i.e. $\hat{x}_{M,\textrm{TV}}=\hat{x}_{0}$)?
\end{itemize}

\subsection{Notations\label{subsec:Notations}}

We firstly introduce some notations that will be used in the rest
of this work. For any complex number $z$, we write by $\bar{z}$
its conjugate. The adjunction of ${\bf X}$ is denoted ${\bf X}^{\herm}$,
wherever ${\bf X}$ is a vector, a matrix, or a linear operator. The
transposition of a matrix or a vector ${\bf X}$ is written ${\bf X}^{\trans}$.
If $P\in\mathbb{C}^{n-1}\left[X\right]$ is a complex polynomial of
the form $P\left(z\right)=\sum_{k=0}^{n-1}p_{k}z^{k}$ then its conjugate
is denoted $P^{\herm}$ and verifies $P^{\herm}\left(z\right)=\sum_{k=0}^{n-1}\bar{p}_{k}z^{k}$
for all $z\in\mathbb{C}$. Unless stated differently, vectors of $\mathbb{C}^{n}$
are indexed in $\left\llbracket 0,n-1\right\rrbracket $ so that every
vector $u\in\mathbb{C}^{n}$ writes $u=\left[u_{0},\dots,u_{n-1}\right]^{\trans}$.
The space of square matrices and the one of Hermitian matrices of
dimension $n$ with complex coefficients are respectively denoted
$\mathrm{M}_{n}\left(\mathbb{C}\right)$ and $\mathrm{S}_{n}\left(\mathbb{C}\right)$.
The cone of positive Hermitian matrices of same dimension is denoted
$\mathrm{S}_{n}^{\splus}\left(\mathbb{C}\right)$. Vectorial spaces
of matrices are all endowed with the Frobenius inner product denoted
$\left\langle \cdot,\cdot\right\rangle $ and defined by $\left\langle A,B\right\rangle =\textrm{tr}\left(A^{\herm}B\right)$,
where $\textrm{tr}\left(\cdot\right)$ is the trace operator. The
canonical Toeplitz Hermitian matrix generator in dimension $n$, denoted
$\mathcal{T}_{n}$, is defined by
\begin{align}
\mathcal{T}_{n}:\;\mathcal{\mathbb{C}}^{n} & \rightarrow\mathrm{M}_{n}\left(\mathbb{C}\right)\nonumber \\
u & \mapsto\mathcal{T}_{n}\left(u\right)=\begin{bmatrix}u_{0} & u_{1} & \dots & u_{n-1}\\
\overline{u_{1}} & u_{0} & \dots & u_{n-2}\\
\vdots & \vdots & \ddots & \vdots\\
\overline{u_{n-1}} & \overline{u_{n-2}} & \dots & u_{0}
\end{bmatrix}.\label{eq:ToeplitzSymmetricGenerator}
\end{align}
Its adjoint $\mathcal{T}_{n}^{*}$ is characterized for every matrix
$H\in\mathrm{M}_{n}\left(\mathbb{C}\right)$ by
\[
\forall k\in\left\llbracket 0,n-1\right\rrbracket ,\quad\mathcal{T}_{n}^{*}\left(H\right)\left[k\right]=\left\langle \Theta_{k},H\right\rangle =\textrm{tr}\left(\Theta_{k}^{\herm}H\right),
\]
whereby $\Theta_{k}$ is the elementary Toeplitz matrix equals to
$1$ on the $k^{\textrm{th}}$ upper diagonal and zero elsewhere,
i.e.
\[
\forall\left(i,j\right)\in\left\llbracket 0,n-1\right\rrbracket ^{2},\quad\Theta_{k}\left(i,j\right)=\begin{cases}
1 & \text{if }j-i=k\\
0 & \text{otherwise}.
\end{cases}
\]
 For every matrix $M\in\mathbb{C}^{m\times n}$, $m\leq n$, we denote
by $\mathcal{R}_{M}$ the operator given by
\begin{align*}
\mathcal{R}_{M}:\;\mathcal{\mathbb{C}}^{n} & \rightarrow\mathrm{M}_{m}\left(\mathbb{C}\right)\\
u & \mapsto\mathcal{R}_{M}\left(u\right)=M\mathcal{T}_{n}\left(u\right)M^{\herm}.
\end{align*}
Its adjoint $\mathcal{R}_{M}^{*}$ is consequently characterized for
every matrix $S\in\mathrm{M}_{m}\left(\mathbb{C}\right)$ by $\mathcal{R}_{M}^{*}\left(S\right)=\mathcal{T}_{n}^{*}\left(M^{\herm}SM\right).$

A selection matrix $C_{\mathcal{I}}\in\left\{ 0,1\right\} ^{m\times n}$
for a subset $\mathcal{I}\subseteq\left\llbracket 0,n-1\right\rrbracket $
of cardinality $m$ is a boolean matrix whose rows are equal to $\left\{ e_{k}^{\herm},k\in\mathcal{I}\right\} $,
where $e_{k}\in\mathbb{C}^{n}$ is the $k^{\textrm{th}}$ vector of
the canonical basis of $\mathbb{C}^{n}$. For a given subset $\mathcal{I}$,
there are $m!$ possible associated sub-sampling matrices, all obtained
by permutation of their rows. For readability, we reduce the respective
notations of the operators $\mathcal{R}_{C_{\mathcal{I}}}$ and $\mathcal{R}_{C_{\mathcal{I}}}^{\herm}$
to $\mathcal{R}_{\mathcal{I}}$ and $\mathcal{R}_{\mathcal{I}}^{\herm}$
for such matrices. 

\subsection{\label{subsec:Preliminaries}Dual problem and certifiability}

It has been shown in \cite{Tang2013} that the primal problem (\ref{eq:GenericPartialL1})
admits for Lagrange dual problem a certain semidefinite program when
the sub-sampling matrix is selection matrix $C_{\mathcal{I}}$. This
result easily extends in our context for any sub-sampling matrix $M$
as stated by the following proposition.
\begin{lem}
[Dual characterization]\label{lem:DualCharacterization}The dual
feasible set $\mathcal{D}_{M}$ of Problem (\ref{eq:GenericPartialL1})
is characterized by
\[
\mathcal{D}_{M}=\left\{ c\in\mathbb{C}^{m},\;\begin{cases}
q=M^{\herm}c\\
\left\Vert Q\left(e^{i2\pi\nu}\right)\right\Vert _{\infty}\leq1
\end{cases}\right\} ,
\]
whereby $Q\in\mathbb{C}^{n-1}\left[X\right]$ is the complex polynomial
having for coefficients vector $q\in\mathbb{C}^{n}$. The Lagrangian
dual of Problem (\ref{eq:GenericPartialL1}) takes the semidefinite
form,
\begin{align}
c_{\star} & =\arg\max_{c\in\mathbb{C}^{m}}\Re\left(y^{\trans}c\right)\label{eq:FullSDP}\\
\textrm{subject to} & \phantom{\phantom{\;=\;}}\begin{bmatrix}H & q\\
q^{\herm} & 1
\end{bmatrix}\succeq0\nonumber \\
\mathcal{} & \phantom{\;=\;}\mathcal{T}_{n}^{*}\left(H\right)=e_{0}\nonumber \\
 & \phantom{\;=\;}q=M^{\herm}c.\nonumber 
\end{align}
\end{lem}
\begin{prop}
[Dual certifiability]\label{prop:DualCertifiability}If there exists
a polynomial $Q_{\star}\in\mathbb{C}^{n-1}\left[X\right]$ having
for coefficients vector $q_{\star}\in\mathbb{C}^{n}$ satisfying the
conditions
\begin{equation}
\begin{cases}
q_{\star}\in\range\left(M^{\herm}\right)\\
Q_{\star}\left(e^{i2\pi\frac{\xi_{r}}{f}}\right)=\sign\left(\alpha_{r}\right), & \forall\xi_{r}\in\Xi\\
\left|Q_{\star}\left(e^{i2\pi\nu}\right)\right|<1, & \text{otherwise},
\end{cases}\label{eq:DualCertificateCondition}
\end{equation}
then the solutions of the Programs (\ref{eq:GenericL0Problem}), and
(\ref{eq:GenericPartialL1}) are unique and one has $\hat{x}_{0}=\hat{x}_{M,\textrm{TV}}$.
Moreover, $\hat{x}=\hat{x}_{M,\textrm{TV}}$ up to an aliasing factor
modulo $f$.
\end{prop}
\begin{IEEEproof}
Any polynomial $Q_{\star}$ satisfying the last two interpolation
conditions of (\ref{eq:DualCertificateCondition}) maximizes the dual
of Problem (\ref{eq:GenericL1Problem}) over the feasible set $D_{I_{n}}$,
and qualifies as a dual certificate of the same problem. Thus, the
solution of Program (\ref{eq:GenericL1Problem}) is unique and satisfies
$\hat{x}_{0}=\hat{x}_{\textrm{TV}}$ \cite{Candes2014a}. By strong
duality, the primal problem (\ref{eq:GenericL1Problem}) and its dual
reach the same optimal objective value, denoted $\kappa_{\star}$. 

By the first condition of (\ref{eq:DualCertificateCondition}), $q_{\star}=M^{\herm}c_{\star}$
for some $c_{\star}\in\mathbb{C}^{m}$. Since $c\in D_{M}\Leftrightarrow M^{\herm}c\in D_{I_{n}}$
for all $c\in\mathbb{C}^{m}$, $c_{\star}$ is dual optimal for the
partial problem $\eqref{eq:GenericPartialL1}$ and reaches the dual
objective $\kappa_{\star}$. By strong duality, $\kappa_{\star}$
also minimize the primal objective of $\eqref{eq:GenericPartialL1}$.
Finally, every feasible point of $\eqref{eq:GenericPartialL1}$ is
feasible for (\ref{eq:GenericL1Problem}). We conclude by uniqueness
of $\hat{x}_{\textrm{TV}}$ on the equality $\hat{x}_{0}=\hat{x}_{\textrm{TV}}=\hat{x}_{M,\textrm{TV}}$.
Finally $\hat{x}=\hat{x}_{0}$ (and thus $\hat{x}_{M,\textrm{TV}}$)
up to an ambiguity modulo $f$ is a direct consequence of Shannon's
sampling theorem.
\end{IEEEproof}
Any polynomial $Q_{\star}$ satisfying the conditions (\ref{eq:DualCertificateCondition})
will be called \emph{dual certificate} for the partial line spectral
estimation problem. Finding meaningful sufficient conditions for the
existence of such dual certificate is a difficult problem in the general
case. One might expect their existence under two main conditions.
The first one comes as a quite intuitive application of the principle
stated in \cite{Tang2015}: the spikes of the signal $\hat{x}$ have
to obey a minimal separability condition of the kind (\ref{eq:MinimalSeparabilityCond})
(for a potentially different constant). The second one is on the sub-sampling
matrix $M$, which has to somehow preserve the spectral properties
of $\hat{x}$, and will be discussed latter.

Sufficient conditions for the existence of a dual certificate will
be detailed in Section \ref{sec:Multirate-sampling-systems} and Section
\ref{sec:Random-sub-sampling} for two different classes of sub-sampling
matrices. Generic results, valid for any arbitrary sub-sampling matrix
$M$, are still lacking and remain an open area of research.

\subsection{Main result}

Lemma \ref{lem:DualCharacterization} proposes to recover the spectral
support of the time signal $x$ by firstly solving a semidefinite
program of dimension $n+1$, and in the latter, to read its output
$c_{\star}$ as a polynomial $Q_{\star}\in\mathbb{C}^{n-1}\left[X\right]$,
where $q_{\star}=M^{\herm}c_{\star}$. The spectral support of $x$
is estimated by the points where this polynomial reaches $1$ in modulus
around the unit circle. However, this method is not satisfactory on
a computational point of view. The complexity of the SDP (\ref{eq:FullSDP})
is driven by the size of its linear matrix inequality, here of size
$n+1$, while the \emph{essential dimension} of partial recovery problem
(\ref{eq:GenericPartialL1}) is equal to the number of measurements
$m\leq n$. In this section, it is shown that, if the matrix $M$
admits a simple admissibility criterion, Program (\ref{eq:FullSDP})
is equivalent to another SDP involving a matrix inequality of lower
dimension equal to $m+1$.
\begin{defn}
[Admissibility condition]\label{def:AdmissibleOperator}A sub-sampling
matrix $M\in\mathbb{C}^{m\times n}$ is said to be \emph{admissible}
if and only if $M$ is full rank and $e_{0}\in\range\left(M^{\herm}\right)$,
where $e_{0}\in\mathcal{\mathbb{C}}^{n}$ is the first vector of the
canonical basis indexed in $\left\llbracket 0,n-1\right\rrbracket $.

Now we are ready to state the main result for this work, whose full
demonstration is provided in Section \ref{sec:Proof-of-Dimension-Reduction}.
\end{defn}
\begin{thm}
[Dimensionality reduction]\label{thm:Dual-CompactSDP-Equivalence}If
the sub-sampling matrix $M\in\mathbb{C}^{m\times n}$ is admissible,
the Lagrange dual problem of Problem (\ref{eq:GenericPartialL1})
is equivalent to the low-dimensional semidefinite program
\begin{align}
c_{\star} & =\arg\max_{c\in\mathbb{C}^{m}}\Re\left(y^{\trans}c\right)\label{eq:ReducedSDP}\\
\textrm{subject to} & \phantom{\phantom{\;=\;}}\begin{bmatrix}S & c\\
c^{\herm} & 1
\end{bmatrix}\succeq0\nonumber \\
 & \phantom{\;=\;}\mathcal{R}_{M}^{*}\left(S\right)=e_{0}.\nonumber 
\end{align}
\end{thm}
A few remarks are in order regarding the statement of Theorem \ref{thm:Dual-CompactSDP-Equivalence}.
First of all, the measurement matrix $M$ has to be admissible for
the theorem to hold. If this condition is not respected, the feasible
set of SDP (\ref{eq:ReducedSDP}) is empty, and obviously differ from
the dual feasible set $\mathcal{D}_{M}$. Secondly, the linear constraint
$\mathcal{R}_{M}^{*}\left(S\right)=e_{0}$ has an explicit dimension
that is still equal to $n$. However, since $M$ is fixed and known,
one can restrict this linear constraint to the span of $\mathcal{R}_{M}^{*}\left(\mathrm{S}_{m}\right)$
which is of dimension lower than $\min\left\{ n,\frac{m\left(m+1\right)}{2}\right\} =\mathcal{O}\left(m^{2}\right)$.
An explicit characterization of this constraint is provided in Section
\ref{subsec:SelectionMatrices} when $M$ is a selection matrix.

\subsection{Case of selection sub-sampling matrices\label{subsec:SelectionMatrices}}

Selection matrices constitutes a particularly interesting type of
sub-sampling matrices, and arise in many practical applications. Their
use is natural in signal processing occur when dealing with sampling
models with missing entries. In this section, we highlight fundamental
properties of the partial line spectrum estimation problem from selection
based sub-sampling. We start by giving a direct characterization of
the admissibility of a matrix $C_{\mathcal{I}}$.
\begin{lem}
\label{lem:SelectionMatrix-Admissibility}A selection matrix $C_{\mathcal{I}}\in\left\{ 0,1\right\} ^{m\times n}$
for a subset $\mathcal{I}\subseteq\left\llbracket 0,n-1\right\rrbracket $
for cardinality $m$ is admissible in the sense of Definition \ref{def:AdmissibleOperator}
if and only if $0\in\mathcal{I}$.
\end{lem}
The proof of the above is trivial and arise directly from the definition
of $C_{\mathcal{I}}$. The next proposition explicits the structure
of $\mathcal{R}_{\mathcal{I}}^{*}$ and recast the linear constraint
$\mathcal{R}_{M}^{*}\left(S\right)=r$ under a more friendly set of
equations.
\begin{prop}
Let $\mathcal{I}\subset\left\llbracket 0,n-1\right\rrbracket $ be
a subset of cardinality $m$ and consider any selection matrix $C_{\mathcal{I}}\in\mathbb{C}^{m\times n}$
for this subset. Define by $\mathcal{J}$ the set of its pairwise
differences $\mathcal{J=\mathcal{I}}-\mathcal{I}$, and by $\mathcal{J}_{+}=\left\{ j\in\mathcal{J},\;j\geq0\right\} $
its positive elements. There exists a skew-symmetric partition of
the square $\left\llbracket 1,m\right\rrbracket ^{2}$ into $p=\left|\mathcal{J}_{+}\right|$
subsets $\left\{ J_{k},\:k\in\mathcal{J}_{+}\right\} $ given by the
support of the matrices $\left\{ C_{\mathcal{I}}^{\herm}\Theta_{k}C_{\mathcal{I}}\right\} _{k\in\mathcal{J}_{\splus}}$
satisfying
\[
\begin{cases}
J_{k}\cap J_{l}=\emptyset, & \forall\left(k,l\right)\in\mathcal{J}_{+}^{2},\;k\neq l,\\
\left(i,j\right)\in\bigcup_{k\in\mathcal{J}_{+}}J_{k}\Leftrightarrow\left(j,i\right)\notin\bigcup_{k\in\mathcal{J}_{+}}J_{k}, & \forall\left(i,j\right)\in\left\llbracket 1,m\right\rrbracket ^{2},\;i\neq j,\\
\left(i,i\right)\in\bigcup_{k\in\mathcal{J}_{+}}J_{k}, & \forall i\in\left\llbracket 1,m\right\rrbracket ,
\end{cases}
\]
 such that, 
\begin{equation}
\forall S\in\mathrm{S}_{m}\left(\mathbb{C}\right),\;\mathcal{R}_{\mathcal{I}}^{*}\left(S\right)=\sum_{k\in\mathcal{J}_{+}}\left(\sum_{\left(l,r\right)\in J_{k}}S_{l,r}\right)e_{k},\label{eq:MkSupportProperty}
\end{equation}
where $e_{k}\in\mbox{\ensuremath{\mathbb{C}}}^{n}$ is the $k^{\textrm{th}}$
vector of the canonical basis of $\mbox{\ensuremath{\mathbb{C}}}^{n}$
indexed in $\left\llbracket 0,n-1\right\rrbracket $.\label{prop:SubspaceConstraint}
\end{prop}
\begin{IEEEproof}
Using the adjoint decomposition of the operator $\mathcal{R}_{\mathcal{I}}^{*}$
on the canonical basis one has,
\begin{align}
\forall S\in\mathrm{S}_{m}\left(\mathbb{C}\right),\quad\mathcal{R}_{\mathcal{I}}^{*}\left(S\right) & =\sum_{k=0}^{n-1}\left\langle \mathcal{R}_{\mathcal{I}}\left(e_{k}\right),S\right\rangle e_{k}\nonumber \\
 & =\sum_{k=0}^{n-1}\left\langle C_{\mathcal{I}}\Theta_{k}C_{\mathcal{I}}^{\herm},S\right\rangle e_{k}.\label{eq:R-Operator analysis}
\end{align}
Let by $M_{k}\in\mathrm{M}_{m}\left(\mathbb{C}\right)$ the matrix
given by $M_{k}=C_{\mathcal{I}}\Theta_{k}C_{\mathcal{I}}^{\herm}$
for all $k\in\left\llbracket 0,n-1\right\rrbracket $. It remains
to show that the support of the matrices $\left\{ M_{k}\right\} _{k\in\left\llbracket 0,n-1\right\rrbracket }$
are forming the desired partition. The general term of matrix $M_{k}$,
obtained by direct calculation, reads
\begin{equation}
\forall\left(i,j\right)\in\left\llbracket 1,m\right\rrbracket ^{2},\quad M_{k}\left(i,j\right)=\begin{cases}
1 & \text{if}\;\mathcal{I}\left[j\right]-\mathcal{I}\left[i\right]=k\\
0 & \text{otherwise},
\end{cases}\label{eq:Mk-GeneralTerm}
\end{equation}
for all $k\in\left\llbracket 0,n-1\right\rrbracket $, whereby $\mathcal{I}\left[j\right]$
represents the $j^{\textrm{th}}$ element of the index set $\mathcal{I}$
for the ordering induced by the matrix $C_{\mathcal{I}}$. The general
term (\ref{eq:Mk-GeneralTerm}) ensures that,
\[
\begin{cases}
M_{0}\left(i,i\right)=1, & \forall i\in\left\llbracket 1,m\right\rrbracket \\
\sum_{k=0}^{n}M_{k}\left(i,j\right)=1\Leftrightarrow\sum_{k=0}^{n}M_{k}\left(j,i\right)=0, & \forall\left(i,j\right)\in\left\llbracket 1,m\right\rrbracket ^{2},\;i\neq j,\\
k\notin\mathcal{J}_{\splus}\Leftrightarrow M_{k}=0_{m}, & \forall k\in\left\llbracket 0,n-1\right\rrbracket ,
\end{cases}
\]
where $0_{m}$ is the null element of $\mathrm{M}_{m}\left(\mathbb{C}\right)$.
Since the matrices $\left\{ M_{k}\right\} _{k\in\left\llbracket 0,n-1\right\rrbracket }$
are constituted of boolean entries, the two first assertions yields
the set of supports $\left\{ J_{k}\right\} _{k\in\left\llbracket 0,n-1\right\rrbracket }$
of $\left\{ M_{k}\right\} _{k\in\left\llbracket 0,n-1\right\rrbracket }$
forms an skew-symmetric partition of $\left\llbracket 1,m\right\rrbracket ^{2}$.
The third one states that only $p=\left|\mathcal{J}_{\splus}\right|$
elements of this partition are non-trivial. After removing those null
matrices, the set $\left\{ J_{k}\right\} _{k\in\mathcal{J}_{+}}$
remains a partition of $\left\llbracket 1,m\right\rrbracket ^{2}$.
We conclude using Equation (\ref{eq:R-Operator analysis}) that,
\begin{align*}
\forall S\in\mathrm{S}_{m}\left(\mathbb{C}\right),\quad\mathcal{R}_{\mathcal{I}}^{*}\left(S\right) & =\sum_{k\in\mathcal{J}_{+}}\left\langle M_{k},S\right\rangle e_{k}\\
 & =\sum_{k\in\mathcal{J}_{+}}\left(\sum_{\left(l,r\right)\in J_{k}}S_{l,r}\right)e_{k}.
\end{align*}
\end{IEEEproof}
This proposition highlights several major properties of the equation
$\mathcal{R}_{\mathcal{I}}^{*}\left(S\right)=r$ for $r\in\mathbb{C}^{n}$:
\begin{itemize}
\item The linear equation is solvable if and only if $r$ is supported in
$\mathcal{J}_{\splus}$, and since $M_{0}=I_{m}$, $r_{0}\in\mathbb{R}$.
\item If so, the equation is equivalent to solve $p=\left|\mathcal{J}_{+}\right|$
linear forms. Those $p$ forms are independent one from the other
in the sense that they are acting on disjoint extractions of the matrix
$S$.
\item The order of each of those forms is smaller that $m$, i.e., each
form involves at most $m$ terms of $S$.
\item The total number of unknowns appearing in this system is exactly $\frac{m\left(m+1\right)}{2}$.
\end{itemize}
In Section \ref{sec:ADMM}, a highly scalable algorithm to solve the
SDP (\ref{eq:ReducedSDP}) for selection matrices, taking advantage
of the hereby presented properties, will be presented.

\section{Multirate sampling systems\label{sec:Multirate-sampling-systems}}

\subsection{Observation model}

A multirate sampling system (MRSS) on a continuous time signal $x$
is defined by a set $\mathbb{A}$ of $p$ distinct grids (or samplers)
$\mathcal{A}_{j}$, $j\in\left\llbracket 1,p\right\rrbracket $. Each
grid is assimilated to a triplet $\mathcal{A}_{j}=\left(f_{j},\gamma_{j},n_{j}\right)$,
where $f_{j}\in\mathbb{R}^{+}$ is its sampling frequency, $\gamma_{j}\in\mathbb{R}$
is its processing delay, expressed in sample unit for normalization
purposes, and $n_{j}\in\mathbb{N}$ the number of measurements acquired
by the grid. We assume those intrinsic characteristics to be known.
The output $y_{j}\in\mathbb{C}^{n_{j}}$ of the grid $\mathcal{A}_{j}$
sampling a complex time signal $x$ following the $s$-spikes model
(\ref{eq:SpikesModel}) reads
\begin{equation}
\forall k\in\left\llbracket 0,n_{j}-1\right\rrbracket ,\quad y_{j}\left[k\right]=\sum_{r=1}^{s}\alpha_{r}e^{i2\pi\frac{\xi_{r}}{f_{j}}\left(k-\gamma_{j}\right)}.\label{eq:ObservationModel}
\end{equation}

Applications of the MRSS framework are numerous in signal processing.
It occurs when sampling in parallel the output of a common channel
in order to get benefits from cleverly designed sampling frequencies
and delays; such design appears, for example, in modern digitalization
with variable bit-rates and analysis of video and audio streams. The
MRSS framework is also naturally fitted to describe sampling processes
in distributed sensor networks: each node, with limited processing
capabilities, samples at its own rate, a delayed version of a complex
signal. Collected data are then sent and merged at a higher level
processing unit, performing a global estimation of the spectral distribution
on a joint manner.

The frequency estimation problem consists, as explained earlier, in
finding the sparsest spectral density that jointly matches the $p$
observation vectors $y_{j}$ for all $j\in\left\llbracket 1,p\right\rrbracket $.
Equivalently to (\ref{eq:GenericL0Problem}), this problem can be
presented by a combinatorial minimization program of the $L_{0}$
pseudo-norm over the set of spectral distributions: 
\begin{align}
\hat{x}_{0} & =\arg\min_{\hat{x}\in D_{1}}\left\Vert \hat{x}\right\Vert _{0}\label{eq:MRSSl0Problem}\\
\text{subject to} & \phantom{\;=\;}y_{j}=\linmeasure_{j}\left(\hat{x}\right),\quad\forall j\in\left\llbracket 1,p\right\rrbracket ,\nonumber 
\end{align}
where $\linmeasure_{j}$ is the linear operator denoting the effect
of the spectral density on the samples acquired by the grid $\mathcal{A}_{j}$
given by
\begin{equation}
\forall j\in\left\llbracket 1,p\right\rrbracket ,\quad\mathcal{L}_{j}=\mathcal{F}_{n,f_{j}}\circ\mathcal{M}_{\frac{\gamma_{j}}{f_{j}}},\label{eq:Def-ObservationOperator}
\end{equation}
whereby the operator $\mathcal{M}_{\tau}$ , $\tau\in\mathbb{R}$
denotes the temporal shift (or spectral modulation) operator defined
for all $h\in D_{1}$ by $\mathcal{M}_{\tau}\left(h\right)\left(\xi\right)=e^{-i2\pi\tau\xi}h\left(\xi\right)$
for all $\xi\in\mathbb{R}$.

Finally, it is important to notice that two different grids $\mathcal{A}_{j}$
and $\mathcal{A}_{j^{\prime}}$ may sample a value of the signal $x$
at the same time instant on the respective sampling indexes $k$ and
$k^{\prime}$, enforcing a relation of the kind $y_{j}\left[k\right]=y_{j^{\prime}}\left[k^{\prime}\right]$.
In the following we denote by $\tilde{m}=$$\sum_{j=1}^{p}n_{j}$
the total number of samples acquired by the system $\mathbb{A}$,
and by $m\leq\tilde{m}$ the net number of observations, obtained
after removing such sampling overlaps, so that $m$ is the number
of \emph{independent observation constraints} of the system. The joint
measurement vector is denoted $\tilde{y}=\left[y_{1}^{\trans},\dots,y_{p}^{\trans}\right]^{\trans}\in\mathbb{C}^{\tilde{m}}$.
We let by $y\in\mathbb{C}^{m}$ its net counterpart by discarding
the redundancies of $\tilde{y}$, so that $y=P_{\mathbb{A}}\tilde{y}$
for some selection matrix $P_{\mathbb{A}}\in\left\{ 0,1\right\} ^{m\times\tilde{m}}$
. The joint linear measurement constraint of Problem (\ref{eq:MRSSl0Problem})
can then be reformulated $y=\mathcal{L}\left(\hat{x}\right)$, where
the operator $\mathcal{L}\in\left(D_{1}\mapsto\mathbb{C}^{m}\right)$
admits the partial operators $\left\{ \linmeasure_{j}\right\} _{j\in\left\llbracket 1,p\right\rrbracket }$
as restrictions on the $p$ subspaces induced by the construction
of the net observation vector $y$.

\subsection{Common grid expansion and SDP formulation\label{subsec:Common-grid-expansion}}

It is been shown in Section \ref{sec:Exact-Dimensionality-Reduction}
that the dual problem can take the form of a low dimensional SDP whenever
the observation operator $\linmeasure$ can be written under the form
$\mathcal{L}=M\mathcal{F}_{n}$ for some measurement matrix $M\in\mathbb{C}^{m\times n}$
satisfying the admissibility condition \ref{def:AdmissibleOperator}.
As highlighted in the proof of Lemma \ref{lem:SelectionMatrix-Admissibility},
this remarkable property is due to the polynomial nature of the adjoint
measurement operator $\mathcal{L}^{*}$. However, in the MRSS context,
the dual observation operator defined by $\linmeasure^{*}\left(c\right)=\sum_{j=1}^{m}\mathcal{\linmeasure}_{j}^{*}\left(c_{j}\right)$
does not take such polynomial form in the general case. A direct calculation
reveals that $\linmeasure^{*}\left(c\right)$ is instead an exponential
polynomial\footnote{A function $f$ of the complex variable $z$ of the form $f\left(z\right)=\sum_{k=1}^{m}c_{k}z^{\gamma_{k}}$
for some $\left\{ \gamma_{k}\right\} _{\left\llbracket 1,m\right\rrbracket }\subset\mathbb{R}$.} for all $c\in\mathbb{C}^{m}$. Up to our knowledge, there is no welcoming
algebraic characterization for optimization purposes of the dual feasible
set $\mathcal{D}_{\mathbb{A}}=\left\{ c\in\mathbb{C}^{m},\left\Vert \mathcal{L}^{\herm}\left(c\right)\right\Vert _{\infty}\leq1\right\} $.
Therefore, the theory developed in Section \ref{sec:Exact-Dimensionality-Reduction}
cannot be directly transcribed in the MRSS framework.

To bridge this concern, we restrict our analysis to the case where
the observation operator admits a factorization of the form $\mathcal{L}=M\mathcal{F}_{n}$
for some $n\in\mathbb{N}$ and $M\in\mathbb{C}^{m\times n}$. The
following aims to provide an algebraic criterion on the parameters
$\left\{ \left(f_{j},\gamma_{j},n_{j}\right)\right\} $ of $\mathbb{A}$
for this hypothesis to hold. We will see that this extra hypothesis
consists in supposing that the samples acquired by $\mathbb{A}$ can
by virtually aligned at a higher rate on another grid $\mathcal{A}_{\splus}$.
Such grid will be called common supporting grid for $\mathbb{A}$,
and are defined as follows.
\begin{defn}
\label{def:CommonSupportingGrid}A grid $\mathcal{A}_{\splus}=\left(f_{\splus},\gamma_{\splus},n_{\splus}\right)$
is said to be a \emph{common supporting grid} for a set of sampling
grids $\mathbb{A}=\left\{ \mathcal{A}_{j}\right\} _{j\in\left\llbracket 1,p\right\rrbracket }$
if and only if the set of samples acquired by the MRSS induced by
$\mathbb{A}$ is a subset of the one acquired by $\mathcal{A}_{\splus}$.
In formal terms, the definition is equivalent to,
\begin{equation}
\left\{ \frac{1}{f_{j}}\left(k_{j}-\gamma_{j}\right),\,j\in\left\llbracket 1,p\right\rrbracket ,\,k_{j}\in\left\llbracket 0,n_{j}-1\right\rrbracket \right\} \subseteq\left\{ \frac{1}{f_{\splus}}\left(k-\gamma_{\splus}\right),\,k\in\left\llbracket 0,n_{\splus}-1\right\rrbracket \right\} .\label{eq:CommonSupportingGrid-Definition}
\end{equation}
The set of common supporting grids of $\mathbb{A}$ is denoted by
$\mathcal{C}\left(\mathbb{A}\right)$. Moreover, a common supporting
grid $\mathcal{A}_{\baro}=\left(f_{\baro},\gamma_{\baro},n_{\baro}\right)$
for $\mathbb{A}$ is said to be \emph{minimal} if and only it satisfies
the minimality condition,
\[
\forall\mathcal{A}_{\splus}\in\mathcal{C}\left(\mathbb{A}\right),\quad n_{\baro}\leq n_{\splus}.
\]
Finally, the \emph{equivalent observation set} of the minimal common
grid $\mathcal{A}_{\baro}$, denoted by $\mathcal{I}$, is the subset
of $\left\llbracket 0,n_{\baro}-1\right\rrbracket $ of cardinality
$m$, formed by the $k$'s for which the time instant $\frac{1}{f_{\baro}}\left(k-\gamma_{\baro}\right)$
is acquired by $\mathbb{A}$.
\end{defn}
\begin{figure}
\centering{}\includegraphics[height=0.15\paperheight]{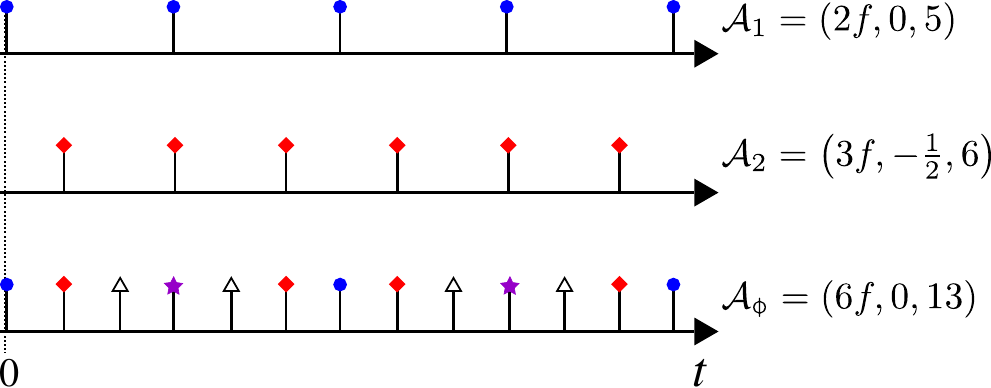}\caption{A representation of a multirate sampling system $\mathbb{A}$ composed
of two arrays $\left(\mathcal{A}_{1},\mathcal{A}_{2}\right)$, and
its associated minimal common grid $\mathcal{A}_{\baro}$. Purple
stars in the common grid correspond to time instant acquired multiple
times by the system $\mathbb{A}$, and blank triangles to omitted
samples. In this example, the dimension of the minimal common grid
is $n_{\baro}=13$, The total number of observation of $\mathbb{A}$,
$\tilde{m}=5+6=11$, and the net number of observations is $m=9$.
Finally the equivalent observation set of the common grid is $\mathcal{I}=\left\{ 0,1,3,5,6,7,9,11,12\right\} $.\label{fig:CommonGrid}}
\end{figure}

It is clear that if $\mathcal{C}\left(\mathbb{A}\right)$ is not empty
then the minimal common supporting grid for $\mathbb{A}$ exists and
is unique. For ease of understanding, Figure \ref{fig:CommonGrid}
illustrates the notion of common supporting grid by showing a MRSS
formed by two arrays and their minimal common grid. Proposition \ref{prop:ExistenceOfCommonSupportingGrid}
states necessary and sufficient conditions in terms of the parameters
of $\mathbb{A}$ such that the set $\mathcal{C}\left(\mathbb{A}\right)$
is not empty. The proof of this proposition is technical and delayed
to Appendix \ref{sec:Proof-Existence of common grid} for readability.
\begin{prop}
\label{prop:ExistenceOfCommonSupportingGrid}Given a set of $p$ grids
$\mathbb{A}=\left\{ \mathcal{A}_{j}=\left(f_{j},\gamma_{j},n_{j}\right)\right\} _{j\in\left\llbracket 1,p\right\rrbracket }$,
the set $\mathcal{C}\left(\mathbb{A}\right)$ is not empty if and
only if there exist $f_{\splus}\in\mathbb{R}^{+}$, $\gamma_{\splus}\in\mathbb{R}$,
a set of $p$ positive integers $\left\{ l_{j}\right\} \in\mathbb{N}^{p}$,
and a set of $p$ integers $\left\{ a_{j}\right\} \in\mathbb{\mathbb{Z}}^{p}$
satisfying $f_{\splus}=l_{j}f_{j}$ and $\gamma_{\splus}=l_{j}\gamma_{j}-a_{j}$
for all $j\in\left\llbracket 1,p\right\rrbracket $. Moreover a common
grid $\mathcal{A}_{\baro}=\left(f_{\baro},\gamma_{\baro},n_{\baro}\right)$
is minimal, if and only if
\[
\begin{cases}
\gcd\left(\left\{ a_{j}\right\} _{j\in\left\llbracket 1,p\right\rrbracket }\cup\left\{ l_{j}\right\} _{j\in\left\llbracket 1,p\right\rrbracket }\right)=1\\
\gamma_{\baro}=\max_{j\in\left\llbracket 1,p\right\rrbracket }\left\{ l_{j}\gamma_{j}\right\} \\
n_{\baro}=\max_{j\in\left\llbracket 1,p\right\rrbracket }\left\{ l_{j}\left(n_{j}-1\right)-a_{j}\right\} .
\end{cases}
\]
\end{prop}
\begin{rem}
\label{rem:CommonGridApproximation}Although the conditions of Proposition
\ref{prop:ExistenceOfCommonSupportingGrid} appear to be strong since
one get $\mathcal{C}\left(\mathbb{A}\right)=\emptyset$ almost surely
in the Lebesgue sense when the sampling frequencies and delays are
drawn at random, assuming the existence of a common supporting grid
for $\mathbb{A}$ is not meaningless in our context. By density, one
can approximately align the system $\mathbb{A}$ on an arbitrary fine
grid $\mathcal{A}_{\varepsilon}$, for any given maximal jitter $\varepsilon>0$,
and perform the proposed super-resolution on this common grid. The
resulting error from this approximation can be interpreted as a ``basis
mismatch''. The detailed analysis of this approach will not be covered
in this work, however, similar approximations can be found in the
literature for the analogue atomic norm minimization view of the super-resolution
problem \cite{Bhaskar2013}. We claim that those results extend in
our settings and that the approximation error vanishes in the noiseless
settings when going to the limit $\varepsilon\rightarrow0$.

The next proposition concludes that the requested factorization of
the linear observation operator $\mathcal{L}$ is possible whenever
$\mathcal{C}\left(\mathbb{A}\right)\neq\emptyset$.
\end{rem}
\begin{prop}
\label{prop:PolynomialAlignment}Let $\mathbb{A}=\left\{ \mathcal{A}_{j}=\left(f_{j},\gamma_{j},n_{j}\right)\right\} _{j\in\left\llbracket 1,p\right\rrbracket }$
be a set of $p$ arrays. The set $\mathcal{C}\left(\mathbb{A}\right)$
is not empty if and only if there exists a subset $\mathcal{I}\subseteq\left[0,n_{\baro}-1\right]$
of cardinality $m$ such that the linear operator $\linmeasure$ defining
the equality constraint of the primal Problem (\ref{eq:GenericL1Problem})
reads,
\[
\linmeasure=C_{\mathcal{I}}\left(\mathcal{F}_{n_{\baro},f_{\baro}}\circ\mathcal{M}_{\frac{\gamma_{\baro}}{f_{\baro}}}\right),
\]
whereby $\mathcal{A}_{\baro}=\left(f_{\baro},\gamma_{\baro},n_{\baro}\right)$
denotes the minimal grid of $\mathbb{A}$ and where $C_{\mathcal{I}}\in\left\{ 0,1\right\} ^{m\times n_{\baro}}$
is a selection matrix of the subset $\mathcal{I}$. Moreover the sub-sampling
matrix $C_{\mathcal{I}}$ is admissible in the sense of Definition
\ref{def:AdmissibleOperator}.
\end{prop}
The proof of this proposition is detailed in Appendix \ref{subsec:Proof-of-PolynomialAlignment}.
The temporal translation $\mathcal{M}_{\frac{\gamma_{\baro}}{f_{\baro}}}$
has little impact in the analysis since a time domain shift leaves
unchanged the spectral support of the probed signal $x$. One can
consider the surrogate signal $x^{\sharp}\left(.\right)=x\left(.-\frac{\gamma_{\baro}}{f_{\baro}}\right)$,
so that $\hat{x}^{\sharp}=\mathcal{M}_{\frac{\gamma_{\baro}}{f_{\baro}}}\left(\hat{x}\right)$
and solve the line spectral estimation problem (\ref{eq:GenericPartialL1})
for the linear constraint $\linmeasure^{\sharp}=C_{\mathcal{I}}\mathcal{F}_{n_{\baro},f_{\baro}}$
via the reduction studied in Section \ref{sec:Exact-Dimensionality-Reduction}.
The complex amplitudes of the spectral $\hat{x}$ can be recover from
its surrogate spectrum by the simple relation $\text{\ensuremath{e^{i2\pi\frac{\gamma_{\baro}}{f_{\baro}}\xi}\alpha^{\sharp}\left(\xi\right)}=\ensuremath{\alpha\left(\xi\right)}}$
for all $\xi\in\mathbb{R}$.

\subsection{Dual certifiability and sub-Nyquist guarantees}

In this section, sufficient conditions are presented to ensure that
the conditions of Proposition \ref{prop:DualCertifiability} are fulfilled.
Those conditions guarantee the tightness of the total-variation relaxation
and the optimality and uniqueness of the recovery $\hat{x}_{0}=\hat{x}_{C_{\mathcal{I}},\textrm{TV}}$.
In addition to this result, it provides mild conditions to ensure
a sub-Nyquist recovery of the spectral spikes at a rate $f_{\baro}$
from measurements taken at various lower rates $\left\{ f_{j}\right\} _{j\in\left\llbracket 1,p\right\rrbracket }$.
The proof of this result, presented in Appendix \ref{subsec:Proof-of-DualCertifiability},
relies on previous polynomial construction methods presented in \cite{Candes2014a,Tang2013,Bhaskar2013}.
\begin{thm}
\label{thm:MRSSDualCertifiability}Let $\mathbb{A}=\left\{ \mathcal{A}_{j}=\left(f_{j},\gamma_{j},n_{j}\right)\right\} _{j\in\left\llbracket 1,p\right\rrbracket }$
be a set of sampling arrays. Suppose that $\mathcal{C}\left(\mathbb{A}\right)$
is not empty, and denote by $\mathcal{A}_{\baro}=\left(f_{\baro},\gamma_{\baro},n_{\baro}\right)$
the minimal common supporting grid of $\mathbb{A}$. Assume that the
system induced by $\mathbb{A}$ satisfies at least one of the two
following separability conditions,
\begin{itemize}
\item \emph{Strong condition:}
\[
\forall j\in\left\llbracket 1,p\right\rrbracket ,\quad\begin{cases}
\Delta_{\mathbb{T}}\left(\frac{1}{f_{j}}\Xi\right)\geq\frac{2.52}{n_{j}-1}\\
n_{j}>2000,
\end{cases}
\]
\item \emph{Weak condition:}
\[
\exists j\in\left\llbracket 1,p\right\rrbracket ,\quad\begin{cases}
\Delta_{\mathbb{T}}\left(\frac{1}{f_{j}}\Xi\right)\geq\frac{2.52}{n_{j}-1}\\
n_{j}>2000\\
m\geq\left(l_{j}+1\right)s,
\end{cases}
\]
\end{itemize}
then there exists a polynomial $Q_{\star}$ verifying the conditions
(\ref{eq:DualCertificateCondition}) Proposition \ref{prop:DualCertifiability}.
Consequently, $\hat{x}_{0}=\hat{x}_{C_{\mathcal{I}},\textrm{TV}}$.
Moreover, $\hat{x}=\hat{x}_{C_{\mathcal{I}},\textrm{TV}}$ up to an
aliasing factor modulo $f_{\baro}$.

\end{thm}
\begin{rem}
First of all, under the weaker proviso $n_{j}>256$, the above results
still hold in both cases when $\Xi$ satisfies the more restrictive
separability criterion $\Delta_{\mathbb{T}}\left(\frac{1}{f_{j}}\Xi\right)\geq\frac{4}{n_{j}-1}$. 

The strong condition for Theorem \ref{thm:MRSSDualCertifiability}
is restrictive and no not particularly highlight any benefits from
jointly estimating the spectral support compared to merging the $p$
spectral estimates obtained by simple individual estimation at each
sampler. However, the weak condition guarantees that frequencies of
the time signal $x$ can be recovered with an ambiguity modulo $f_{\baro}$
when jointly resolving the MRSS, while individual estimations would
guarantee to recover them with an ambiguity modulo $f_{j}\leq f_{\baro}$.
The weak condition require standard spectral separation from a single
array $\mathcal{A}_{j}$, and sufficient net measurements $m$ of
the time signal. The extra measurements $m-n_{j}$ corresponding to
the other grids are\emph{ not uniformly aligned} with the sampler
$\mathcal{A}_{j}$. Therefore the sampling system induced by $\mathbb{A}$
achieves sub-Nyquist spectral recovery of the spectral spikes, and
pushes away the classic spectral range $f_{j}$ from a factor $\frac{f_{\baro}}{f_{j}}=l_{j}$.
Nevertheless, the provided construction of the dual certificate results
in a polynomial having a modulus close to unity on the aliasing frequencies
induced by the zero forcing upscaling from $f_{j}$ to $f_{\baro}$.
Consequently, one can expect to obtain degraded performances in noisy
environments when the sub-sampling factor $l_{j}$ becomes large.
\end{rem}

\subsection{Benefits of multirate measurements\label{subsec:Benefits-of-multirate-measurements}}

Multirate sampling has been applied in many problematics arising from
signal processing and telecommunications in order to reduce either
the number of required measurements or the processing complexity \cite{5609222}.
There are three major benefits of making use of MRSS acquisition in
the line spectral estimation problem. One might just think MRSS has
an obvious way of increasing the number of samples acquired by system
compared to a single grid measurement $\mathcal{A}_{j}\in\mathbb{A}$.
This naturally leads to an enhanced \emph{noise robustness}. More
importantly, MRSS acquisition brings benefits in terms \emph{spectral
range} extension, and \emph{spectral} \emph{resolution} improvement.
The spectral range extension (or sub-Nyquist) capabilities have been
described in Theorem \ref{thm:Dual-CompactSDP-Equivalence}. The spectral
resolution \textemdash{} the minimal distance on the torus between
two spectral spikes to guarantee their recovery \textemdash , is also
expected to be enhanced in MRSS acquisition due to the observation
of delayed versions of the time signal $x$, which virtually enlarges
the global observation window. The resolution guarantees in MRSS will
not be covered in this work and are left for future research.

\begin{figure}
\centering{}\includegraphics[width=0.45\columnwidth]{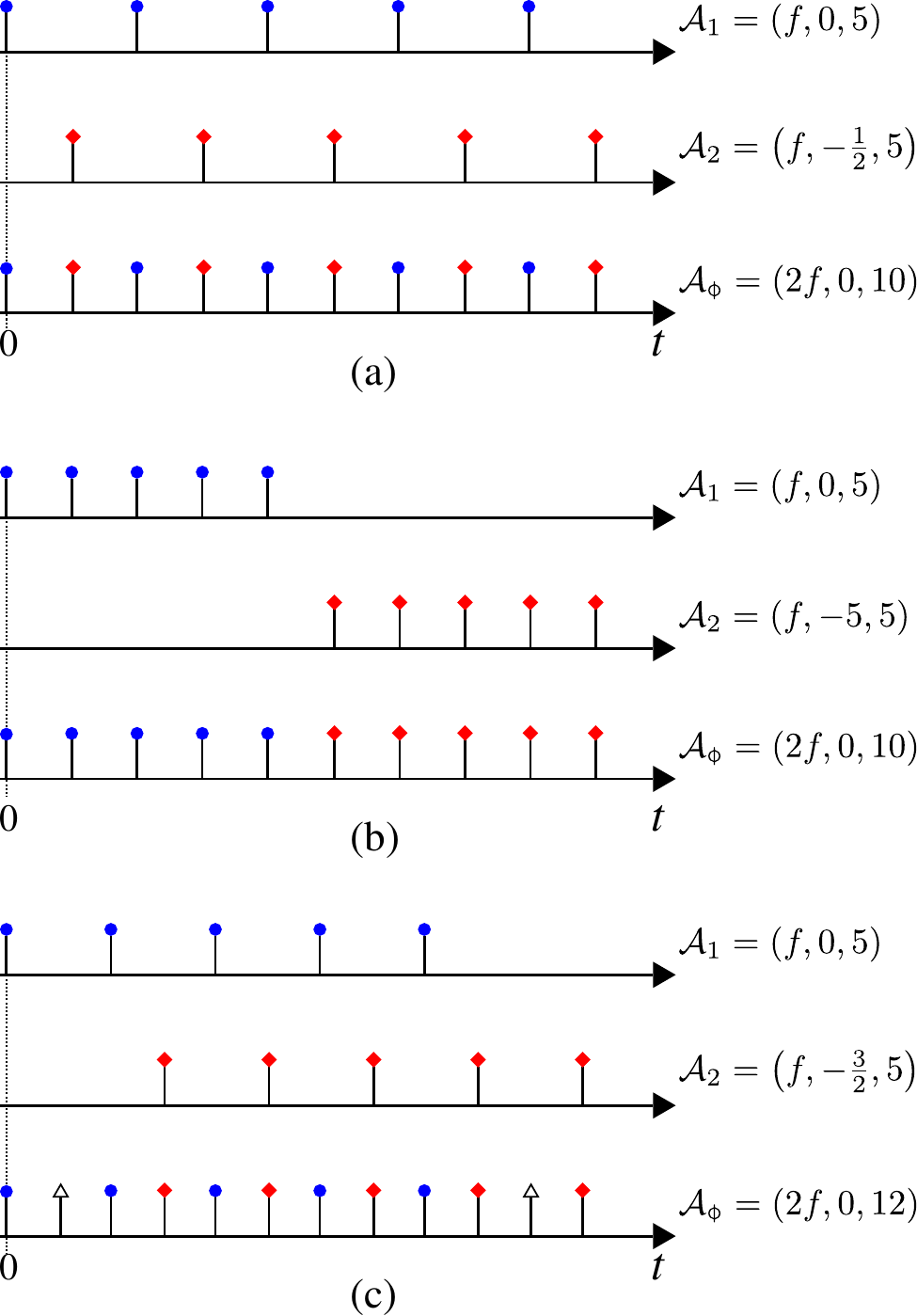}\caption{\label{fig:BenefitOfMRSS}A representation of three delay-only MRSS
in different remarkable settings. In (a), the delay between the two
samplers is exactly of half-unit, resulting in a doubled frequency
range in the joint analysis. In (b), this delay is such that the overall
process equivalently acquires samples on a doubled time frame, resulting
in a doubled spectral resolution. Sub-figure (c) represents an hybrid
case where both resolution improvement and spectral range extension
are expected.}
\end{figure}

For the sake of clarity, Figure \ref{fig:BenefitOfMRSS} proposes
a comprehensive illustration of the trade-off between range extension
and resolution improvement for a delay-only MRSS constituted of two
samplers $\mathcal{A}_{1}$ and $\mathcal{A}_{2}$. In Figure \ref{fig:BenefitOfMRSS}
(a), the delay between the two samplers is such that the joint uniform
grid $\mathcal{A}_{\baro}$ has no missing observations with a double
sampling frequency. One trivially expects to recover the spikes location
of $x$ with aliasing ambiguity modulo $2f$. In \ref{fig:BenefitOfMRSS}(b),
the delay of $\mathcal{A}_{2}$ is set such that the resulting minimal
common grid has a doubled observation window. $\mathcal{A}_{\baro}$
fits again in the uniform observation framework analyzed in \cite{Candes2014a},
and the sufficient spectral separation from the joint measurements
is twice smaller than for the single estimation case. Finally a hybrid
case is presented in Figure \ref{fig:BenefitOfMRSS}(c), where one
expect to get some spectral range and resolution improvements from
a joint recovery approach.

\subsection{Complexity improvements\label{subsec:The-complexity-concern}}

Proposition \ref{prop:PolynomialAlignment} states that, under the
existence of a common grid, the selection operator $C_{\mathcal{I}}\in\left\{ 0,1\right\} ^{m\times n_{\baro}}$
is admissible, consequently Theorem \ref{thm:Dual-CompactSDP-Equivalence}
apply and the dual line spectrum estimation problem can be formulated,
in the MRSS context, by an SDP of dimension $m+1$. In this section,
we highlight the important impact in term of complexity in the MRSS
case.

The original semidefinite program (\ref{eq:FullSDP}) involves a linear
matrix inequality of dimension of $n_{\baro}+1$. The actual value
of $n_{\baro}$, fully determined of the observation pattern induced
by $\mathbb{A}$, reads
\[
n_{\baro}=\max_{j\in\left\llbracket 1,p\right\rrbracket }\left\{ l_{j}\left(n_{j}-1\right)-a_{j}\right\} ,
\]
whereby the parameters $\left\{ \left(a_{j},l_{j}\right)\right\} _{j\in\left\llbracket 1,p\right\rrbracket }$
are defined in Proposition \ref{prop:ExistenceOfCommonSupportingGrid}.
This is particularly disappointing since $n_{\baro}$ grows at a speed
driven by the product of the $n_{j}$'s, whereas the \emph{essential
dimension} $m$ of the problem is given by the number of net observations
acquired by the grid $m\leq\tilde{m}=\sum_{j=1}^{p}n_{j}$. We study
the asymptotic ratio $\frac{m}{n_{\baro}}$ when the number grids
$p$ grows large in two different idealized instances of MRSS to illustrate
that the reduced SDP formulation (\ref{eq:ReducedSDP}) brings \emph{orders
of magnitude} changes to the computational complexity of the line
spectral estimation problem.

Suppose a delay-only MRSS, where $\mathbb{A}$ is constituted of $p$
grids given by $\mathcal{A}_{1}=\left(f,0,n_{0}\right)$ and $\mathcal{A}_{j}=\left(f,-\frac{1}{b_{j}},n_{0}\right)$
for all $j\in\left\llbracket 2,p\right\rrbracket $. Moreover suppose
that $\left\{ b_{j}\right\} _{j\in\left\llbracket 2,p\right\rrbracket }$
are jointly coprime. It is easy to verify the $\mathcal{C}\left(\mathbb{A}\right)$
is not empty in those settings, and that the minimal common grid $\mathcal{A}_{\baro}$
is given by $\mathcal{A}_{\baro}=\left(\left(\prod_{j=2}^{p}b_{j}\right)f,0,\left(\prod_{j=2}^{p}b_{j}\right)n_{0}\right)$.
One has $n_{\baro}=\Omega\left(b^{p}n_{0}\right)$ for some constant
$b\in\mathbb{R}^{+}$, while $m=pn_{0}$. The ratio $\frac{m}{n_{\baro}}=o\left(\frac{p}{b^{p}}\right)$
and tends to $0$ exponentially fast with the number of samplers $m$
of the system. 

On the other hand, suppose a synchronous coprime sampling system between
the time instants $0$ and $T$, where $\mathcal{A}_{j}=\left(k_{j}f,0,k_{j}fT\right)$
for all $j\in\left\llbracket 1,p\right\rrbracket $ with $\gcd\left\{ k_{j},j\in\left\llbracket 1,p\right\rrbracket \right\} =1$.
Once again $\mathcal{C}\left(\mathbb{A}\right)$ is not empty, and
the minimal grid is characterized by the parameters $\mathcal{A}_{\baro}=\left(\left(\prod_{j=1}^{p}k_{j}\right)f,0,\left(\prod_{j=1}^{p}k_{j}\right)fT\right)$.
Consequently $\frac{m}{n_{\baro}}=\frac{\sum_{j=1}^{p}k_{j}}{\prod_{j=1}^{p}k_{j}}$
deceases in $o\left(k^{-p}\right)$ for judicious choice of $\left\{ k_{j}\right\} _{j\in\left\llbracket 1,p\right\rrbracket }$.

\section{Random selection sampling\label{sec:Random-sub-sampling}}

\subsection{Observation model and previous results\label{subsec:RandomSubSampling-ObsevationModel}}

In this section, we consider the line spectrum estimation problem
from a category of selection matrix $C_{\mathcal{I}}\in\left\{ 0,1\right\} ^{m\times n}$
obtained by randomly selecting the observation subset $\mathcal{I}.$
This problem has been introduced in \cite{Tang2013}, and sufficient
conditions to guarantee the tightness of Program (\ref{eq:GenericPartialL1})
have been provided. We hereby summarize those results and introduce
our low dimensional approach to recover the frequencies of a sparse
signal $x$ in those measurement settings.

The observation subset $\mathcal{I}\subseteq\left\llbracket 0,n-1\right\rrbracket $
is constructed by keeping at random, and independently from the others,
each of the elements of $\left\llbracket 0,n-1\right\rrbracket $
with probability $p$, and discarding the rest of it. As a result,
$\mathcal{I}$ has an expected cardinality $\bar{m}=\mathbb{E}\left[\left|\mathcal{I}\right|\right]=pn$.
We consider a subset $\mathcal{I}$, of cardinality $m$ resulting
from the described stochastic process, and recall the following result
from \cite[Theorem I.1]{Tang2013}.
\begin{thm}
[Tang, Bhaskar, Shah, Recht '12]\label{thm:RandomSubSampling-Certifiability}Consider
the partial observation problem (\ref{eq:GenericPartialL1}) with
a sub-sampling matrix $M=C_{\mathcal{I}}\in\left\{ 0,1\right\} ^{m\times n}$
drawn according to the random selection sampling model. Suppose that
the observed signal $x$ following model (\ref{eq:SpikesModel}) satisfies
the spectral separability condition $\Delta_{\mathbb{T}}\left(\frac{1}{f}\Xi\right)\geq\frac{4}{n-1}.$
Moreover, suppose that the phases of the complex amplitudes $\left\{ \alpha_{r}\right\} _{r\in\left\llbracket 1,s\right\rrbracket }$
characterizing the signal $x$ are drawn independently and uniformly
at random in $\left[0,2\pi\right)$. Consider any positive number
$\delta>0$. There exists a constant $C>0$ such that if
\[
m\geq C\max\left\{ \log^{2}\frac{n}{\delta},s\log\frac{s}{\delta}\log\frac{n}{\delta}\right\} ,
\]
then there exists, with probability greater than $1-\delta$, a polynomial
$Q_{\star}$ verifying the conditions (\ref{eq:DualCertificateCondition}).

Consequently, the output of the relaxed Problem (\ref{eq:GenericPartialL1})
is unique and verifies $\hat{x}_{0}=\hat{x}_{C_{\mathcal{I}},\textrm{TV}}$.
Moreover, $\hat{x}=\hat{x}_{C_{\mathcal{I}},\textrm{TV}}$ up to an
aliasing factor modulo $f$.
\end{thm}

\subsection{Dimensionality reduction}

The dimensionality reduction result presented in Theorem \ref{thm:Dual-CompactSDP-Equivalence}
requires an admissible selection matrix $C_{\mathcal{I}}$, i.e. that
$0\in\mathcal{I}$. In the latter, we show that we can always fall
back into this case via some simple considerations similar to the
one described in Section \ref{subsec:Common-grid-expansion}. Let
$k_{0}=\min\mathcal{I}$, and let $x^{\sharp}\left(\cdot\right)=x\left(\cdot-\frac{k_{0}}{f}\right)$.
In the spectral domain the definition reads $\hat{x}^{\sharp}=\mathcal{M}_{\frac{k_{0}}{f}}\hat{x}$,
and the line spectral estimation problem can be equivalently solved
for the spectral density $\hat{x}^{\sharp}$ for the measurement constraint
\[
\linmeasure^{\sharp}=C_{\mathcal{I}-k_{0}}\mathcal{F}_{n}.
\]
One has $0\in\mathcal{I}-k_{0}$ and the selection matrix $C_{\mathcal{I}-k_{0}}\in\mathbb{C}^{m\times n}$
is thus admissible in the sense of Definition \ref{def:AdmissibleOperator}.
It is therefore possible to recover $\hat{x}^{\sharp}$ from the reduced
SDP (\ref{eq:ReducedSDP}), and to reconstruct in a second time $\hat{x}$
via the simple phase shift $\alpha\left(\xi\right)=e^{-i2\pi\frac{\gamma_{\baro}}{f_{\baro}}\xi}\text{\ensuremath{\alpha^{\sharp}\left(\xi\right)}}$
. We are now allowed to conclude on the following result.
\begin{cor}
Under the same hypothesis than Theorem \ref{thm:RandomSubSampling-Certifiability},
the reduced SDP (\ref{eq:ReducedSDP}) of dimension $m+1$ outputs
a polynomial $Q_{\star}$ verifying the conditions (\ref{eq:DualCertificateCondition}).
\end{cor}
Theorem \ref{thm:RandomSubSampling-Certifiability} guarantees a high-probability
recovery of supporting frequencies of the probed signal whenever the
number of measurement grow essentially as the logarithm of $n$. Therefore
using the reduced SDP (\ref{eq:ReducedSDP}) to solve the line spectral
estimation problem brings again \emph{orders of magnitude} changes
in term of computational complexity. The complexity is lowered from
solving the SDP\& of dimension $n$, to solving a SDP having a \emph{poly-logarithmic
}dimension dependency $m=\mathcal{O}\left(\max\left\{ \log^{2}\frac{n}{\delta},s\log\frac{s}{\delta}\log\frac{n}{\delta}\right\} \right)$.

\section{Spectral estimation in noise\label{sec:SpectralEstimationInNoise} }

Up to here, only the case of noise-free spectral estimation has been
studied. In this part, we consider partial noisy observations of a
sparse signal $x$ following the spikes model given in (\ref{eq:SpikesModel})
under the form
\[
\begin{cases}
y_{\mathrm{raw}}\left[k\right]=\sum_{r=1}^{s}\alpha_{r}e^{i2\pi\frac{\xi_{r}}{f}k}+w\left[k\right], & \forall k\in\left\llbracket 0,n-1\right\rrbracket \\
y=My_{\mathrm{raw}},
\end{cases}
\]
for some sub-sampling matrix $M\in\mathbb{C}^{m\times n}$. The noise
vector $w\in\mathbb{C}^{n}$ is assumed to be drawn according to the
spherical $n$ dimensional complex Gaussian distribution $\mathcal{N}\left(0,\sigma^{2}I_{n}\right)$.
We introduce an adapted version of the original Atomic Soft Thresholding
(AST) method, introduced in \cite{Tang2013a} to denoise the spectrum
of $x$ and attempt to retrieve the set of frequencies $\Xi$ supporting
the spectral spikes. The AST method is reviewed to perform in the
partial observation context. Its Lagrange dual version is introduced,
and benefits from the same dimensionality reduction properties than
discussed in Section \ref{sec:Exact-Dimensionality-Reduction}. The
Primal-AST problem consists in optimizing the cost function

\begin{equation}
\hat{x}_{\mathrm{TV}}=\arg\min_{\hat{x}\in D_{1}}\left\Vert \hat{x}\right\Vert _{\mathrm{TV}}+\frac{\tau}{2}\left\Vert y-M\mathcal{F}_{n}\left(\hat{x}\right)\right\Vert _{2}^{2},\label{eq:PrimalAST}
\end{equation}
whereby $\tau\geq0$ is a regularization parameter trading between
the sparsity of the recovered spectrum and the denoising power. Making
use of Proposition \ref{prop:SubspaceConstraint}, if $M$ satisfies
the admissibility condition given in Definition \ref{def:AdmissibleOperator},
the Dual-AST problem is equivalent to the low-dimensional semidefinite
program

\begin{align}
c_{\star} & =\arg\max_{c\in\mathbb{C}^{m}}\Re\left(y^{\trans}c\right)-\frac{\tau}{2}\left\Vert c\right\Vert _{2}^{2}\label{eq:DualAST}\\
\text{subject to} & \phantom{\phantom{\;=\;}}\begin{bmatrix}S & c\\
c^{\herm} & 1
\end{bmatrix}\succeq0\nonumber \\
 & \phantom{\;=\;}\mathcal{R}_{M}^{*}\left(S\right)=e_{0}.\nonumber 
\end{align}
Slatter's condition holds once again for Problem (\ref{eq:PrimalAST}),
and strong duality between (\ref{eq:PrimalAST}) and (\ref{eq:DualAST})
is ensured. Applying the results in \cite{Bhaskar2013}, the choice
of regularization parameter $\tau=\gamma\sigma\sqrt{m\log m}$, for
some $\gamma>1$, is suitable to guarantee a perfect asymptotic recovery
of the spectral distribution $\hat{x}$, while providing accelerated
rates of convergence.

\section{\label{sec:ADMM}Estimation via alternating direction method of multipliers}

\subsection{Interior point methods and ADMM}

Computing the solution of semidefinite program using out of the box
SDP solvers such as \noun{SuDeMi} \cite{Sturm1999} or \noun{SDPT3
\cite{doi:10.1080/10556789908805762}} requires at most $\mathcal{O}\left(\left(m_{\textrm{lmi}}^{2}+m_{\textrm{lin}}\right)^{3.5}\right)$
operations where $m_{\textrm{lmi}}$ is the dimension of the linear
matrix inequality, and $m_{\textrm{lin}}$ the dimension of the linear
constraints. For the dual-AST program (\ref{eq:DualAST}), $m_{\textrm{lmi}}=m+1$
and $m_{\textrm{lin}}\leq\frac{m\left(m+1\right)}{2}$, and approaching
the optimal dual solution will cost $\mathcal{O}\left(m^{7}\right)$
operations using those interior point methods. It appears to be unrealistic
to recover the sparse line spectrum of $x$ that way when the number
of observations exceeds a few hundreds.

In the same spirit than in \cite{Tang2013a}, we derive the steps
and update equations to approach the optimal solution via the alternating
direction method of multipliers (ADMM). Unlike the original work,
we choose to perform ADMM on the dual space instead of the primal
one, and adjust the update steps in order to take advantage of the
low dimensionality of (\ref{eq:DualAST}). The overall idea of this
algorithm is to cut the augmented Lagrangian of the problem into a
sum of separable sub-functions. Each iteration consists in performing
independent local minimization on each of those quantities. The interested
reader can find a detailed survey of this method in \cite{Boyd2010}.

We restrict our analysis to the case of partially observed systems
where the sub-matrix is a selection matrix $C_{\mathcal{I}}\in\left\{ 0,1\right\} ^{m\times n}$
for some subset $\mathcal{I}\subseteq\left\llbracket 0,n-1\right\rrbracket $
of cardinality $m$. We will see that the properties of such matrices
detailed in Section \ref{subsec:SelectionMatrices} will help breaking
down the iterative steps of dual ADMM on an elegant manner. Before
any further analysis, the Dual-AST (\ref{eq:DualAST}) has to be restated
into a more friendly form to derive the ADMM update equations. In
our approach, we propose the following augmented formulation
\begin{align}
c_{\star} & =\arg\min_{c\in\mathbb{C}^{m}}-\Re\left(y^{\trans}c\right)+\frac{\tau}{2}\left\Vert c\right\Vert _{2}^{2}\label{eq:ReducedSDP-ADMM form}\\
\text{subject to} & \phantom{\phantom{\;=\;}}Z\succeq0\nonumber \\
 & \phantom{\phantom{\;=\;}}Z=\begin{bmatrix}S & c\\
c^{\herm} & 1
\end{bmatrix}\nonumber \\
 & \phantom{\;=\;}\sum_{\left(i,j\right)\in J_{k}}S_{i,j}=\delta_{k},\quad k\in\mathcal{J}_{+},\nonumber 
\end{align}
whereby $\delta_{k}$ is the Kronecker symbol. It is immediate, using
Proposition \ref{prop:SubspaceConstraint}, to verify that Problems
(\ref{eq:DualAST}) and (\ref{eq:ReducedSDP-ADMM form}) are actually
equivalent.

\subsection{Lagrangian separability}

We denote by $L$ the restricted Lagrangian of the Problem (\ref{eq:ReducedSDP-ADMM form}),
obtained by ignoring the semidefinite constraint $Z\succeq0$. In
order to ensure plain differentiability with respect to the variables
$S$ and $Z$, ADMM seeks to minimize an augmented version $L_{+}$
of $L$, with respect to the semidefinite inequality constraint that
was put apart. This augmented Lagrangian $L_{+}$ is introduced as
follows
\[
L_{+}\left(Z,S,c,\Lambda,\mu\right)=L\left(Z,S,c,\Lambda,\mu\right)+\frac{\rho}{2}\left\Vert Z-\begin{bmatrix}S & c\\
c^{\herm} & 1
\end{bmatrix}\right\Vert _{F}^{2}+\frac{\rho}{2}\sum_{k\in\mathcal{J}_{+}}\left(\sum_{\left(i,j\right)\in J_{k}}S_{i,j}-\delta_{k}\right)^{2},
\]
whereby the variables $\Lambda\in\mathrm{S}_{m+1}\left(\mathbb{C}\right)$
and $\mu\in\mathbb{C}^{\left|\mathcal{J}_{+}\right|}$ denote respectively
the Lagrange multipliers associated with the first and the second
equality constraints of Problem (\ref{eq:ReducedSDP-ADMM form}).
The regularizing parameter $\rho>0$ is set to ensure a well conditioned
differentiability and to fasten the convergence speed of the alternating
minimization towards the global optimum of the cost function $L_{\splus}$.
For clarity and convenience, the following decompositions of the parameters
$Z$ and $\Lambda$ are introduced
\[
Z=\begin{bmatrix}Z_{0} & z\\
z^{\herm} & \zeta
\end{bmatrix}\qquad\Lambda=\begin{bmatrix}\Lambda_{0} & \lambda\\
\lambda^{\herm} & \eta
\end{bmatrix}.
\]
Moreover, for any square matrix $A\in\mathrm{M}_{m}\left(\mathbb{C}\right)$,
we let by $A_{J_{k}}\in\mathbb{C}^{\left|J_{k}\right|}$ the vector
constituted of the terms $\left\{ A_{i,j},\;\left(i,j\right)\in J_{k}\right\} $.
The order in which the elements of $J_{k}$ are extracted and placed
in this vector has no importance, as long as, once chosen, it remains
the same for every matrix $A$. This allows to decompose the augmented
Lagrangian into
\[
L_{+}\left(Z,S,c,\Lambda,\mu\right)=L_{c}\left(z,c,\lambda\right)+L_{\gamma}\left(\zeta,\eta\right)+\sum_{k\in\mathcal{J_{\splus}}}L_{k}\left(Z_{0,J_{k}},S_{J_{k}},\Lambda_{0,J_{k}}\right),
\]
whereby each of the sub-functions reads
\begin{align*}
L_{c}\left(z,c,\lambda\right) & =-\Re\left(y^{\trans}c\right)+\frac{\tau}{2}\left\Vert c\right\Vert _{2}^{2}+2\left\langle \lambda,z-c\right\rangle +\rho\left\Vert z-c\right\Vert _{2}^{2}\\
L_{\gamma}\left(\zeta,\eta\right) & =\left\langle \eta,\zeta-1\right\rangle +\frac{\rho}{2}\left(\zeta-1\right)^{2}\\
\forall k\in\mathcal{J_{\splus}},\quad L_{k}\left(Z_{0,J_{k}},S_{J_{k}},\Lambda_{0,J_{k}}\right) & =\left\langle \Lambda_{0,J_{k}},Z_{0,J_{k}}-S_{J_{k}}\right\rangle +\mu_{k}\left(\sum_{\left(i,j\right)\in J_{k}}S_{i,j}-\delta_{k}\right)\\
 & \phantom{=\;}+\frac{\rho}{2}\left\Vert Z_{0,J_{k}}-S_{J_{k}}\right\Vert _{2}^{2}+\frac{\rho}{2}\left(\sum_{\left(i,j\right)\in J_{k}}S_{i,j}-\delta_{k}\right)^{2}.
\end{align*}

\subsection{Update rules}

The ADMM will consist in successively performing the following decoupled
update steps:
\begin{align*}
c^{t+1} & \leftarrow\arg\min_{c}L_{c}\left(z^{t},c,\lambda^{t}\right)\\
\forall k\in\mathcal{J}_{+},\quad S_{J_{k}}^{t+1} & \leftarrow\arg\min_{S_{J_{k}}}L_{k}\left(Z_{0,J_{k}}^{t},S_{J_{k}},\Lambda_{0,J_{k}}^{t}\right)\\
S_{j,i}^{t+1} & \leftarrow\overline{S_{i,j}^{t+1}},\hspace{1em}\forall\left(i,j\right)\in\bigcup_{k\in\mathcal{J}_{+}}J_{k}\\
Z^{t+1} & \leftarrow\arg\min_{Z\succeq0}L_{\splus}\left(Z,S^{t+1},c^{t+1},\Lambda^{t},\mu^{t}\right)\\
\Lambda^{t+1} & \leftarrow\Lambda^{t}+\rho\left(Z^{t+1}-\begin{bmatrix}S^{t+1} & c^{t+1}\\
c^{t+1^{\herm}} & 1
\end{bmatrix}\right)\\
\forall k\in\mathcal{J}_{+},\quad\mu^{t+1}\left(k\right) & \leftarrow\mu^{t}\left(k\right)+\rho\left(\sum_{\left(i,j\right)\in J_{k}}S_{i,j}^{t+1}-\delta_{k}\right).
\end{align*}
Since the linear constraint $\mathcal{R}_{\mathcal{I}}^{*}\left(S\right)=e_{0}$
has an effect limited to the subspace $\left\{ \mathcal{R}_{\mathcal{I}}\left(e_{k}\right)\right\} _{k\in\mathcal{J}_{\splus}}$,
the third update step is necessary to maintain the Hermitian structure
of the matrix $S^{t+1}$ at every iteration. The update steps for
the variables $c^{t+1}$ and $\left\{ S_{J_{k}}^{t+1}\right\} _{k\in\mathcal{J}_{+}}$
are performed at each iteration by canceling the gradient of their
partial augmented Lagrangian and admit, in the presented settings,
closed form expressions given by

\begin{align*}
c^{t+1} & =\frac{1}{2\rho+\tau}\left(\bar{y}+2\rho z^{t}+2\lambda^{t}\right)\\
\forall k\in\mathcal{J}_{+},\quad S_{J_{k}}^{t+1} & =\left(Z_{0}^{t}+\frac{1}{\rho}\Lambda_{0}^{t}\right)_{J_{k}}-\left(\sum_{\left(i,j\right)\in J_{k}}\left(Z_{0}^{t}+\frac{\Lambda_{0}^{t}}{\rho}\right)_{i,j}-\left(\delta_{k}-\frac{\mu_{k}^{t}}{\rho}\right)\right)j_{\left|J_{k}\right|}
\end{align*}
whereby $\bar{y}\in\mathbb{C}^{m}$ denotes the conjugate of the observation
vector $y$, and $j_{v}$ is the all-one vector of $\mathbb{C}^{v}$
for all $v\in\mathbb{N}$. The update $Z^{t+1}$ reads at the $t^{\mathrm{th}}$
iteration
\begin{align*}
Z^{t+1} & \in\arg\min_{Z\succeq0}\left\Vert Z-Y^{t}\right\Vert _{F}^{2}\\
Y^{t} & =\begin{bmatrix}S^{t+1} & c^{t+1}\\
c^{t+1^{\herm}} & 1
\end{bmatrix}-\frac{\Lambda^{t}}{\rho},
\end{align*}
which can be interpreted as an orthogonal projection of $Y^{t}$ onto
$\mathrm{S}_{m+1}^{+}\left(\mathbb{C}\right)$ for the Frobenius inner
product. This projection can be computed by looking for the eigenpairs
of $Y^{t}$, and setting all negative eigenvalues to $0$. More precisely,
denoting $Y^{t}=V^{t}D^{t}V^{t^{\herm}}$ an eigen-decomposition of
$Y^{t}$, one get $Z^{t+1}=$$V^{t}D_{+}^{t}V^{t^{\herm}}$ where
$D_{+}^{t}$ is a diagonal matrix whose $j^{\textrm{th}}$ diagonal
entry $d_{\splus}^{t}\left[j\right]$ satisfies $d_{\splus}^{t}\left[j\right]=\max\left\{ d^{t}\left[j\right],0\right\} $.

\subsection{Computational complexity}

On the computational point of view, at each step of ADMM, the update
$c^{t+1}$ is a vector addition and performed in a linear time $\mathcal{O}\left(m\right)$.
On every extractions $S_{J_{k}}^{t+1}$ of $S^{t+1}$, the update
equation is assimilated to a vector averaging requiring $\mathcal{O}\left(\left|J_{k}\right|\right)$
operations when firstly calculating the common second term of the
addition. Since $\bigcup_{k\in\mathcal{J}_{+}}J_{k}=\frac{m\left(m+1\right)}{2}$,
we conclude that the global update of the matrix $S^{t+1}$ is done
in $\mathcal{O}\left(m^{2}\right)$. The update of $Z^{t+1}$ requires
the computation of its spectrum, which can be done in $\mathcal{O}\left(m^{3}\right)$
via power method. Finally updating the multipliers $\Lambda^{t+1}$
and $\mu^{t+1}$ consist in simple matrix and vector additions, thus
of order $\mathcal{O}\left(m^{2}\right)$.

To summarize, the projection is the most costly operation of the loop.
Each step of ADMM method runs in $\mathcal{O}\left(m^{3}\right)$
operations, which is a significant improvement compared to the infeasible
path approached used by SDP solvers requiring around $\mathcal{O}\left(m^{7}\right)$
operations.

\section{Proof of Theorem \ref{thm:Dual-CompactSDP-Equivalence}\label{sec:Proof-of-Dimension-Reduction}}

\subsection{Gram parametrization of trigonometric polynomials}

We start the demonstration by introducing a couple of notations and
by a brief review of the Gram parametrization theory of trigonometric
polynomials. For every non-zero complex number $z\in\mathbb{C}^{*}$,
its $n$-length power vector $\psi_{n}\left(z\right)\in\mathbb{C}^{n}$
is defined by $\psi_{n}\left(z\right)=\left[1,z,\dots,z^{n-1}\right]^{\trans}$.
A complex trigonometric polynomial $R\in\mathbb{C}^{\bar{n}}\left[X\right]$
of order $\bar{n}=2n-1$ is a linear combination of complex monomials
with positive and negative exponents absolutely bounded by $n$. Such
polynomial $R$ reads

\[
\forall z\in\mathbb{C}^{*},\quad R\left(z\right)=\sum_{k=-n+1}^{n-1}r_{k}z^{k}.
\]
It is easy to verify that a complex trigonometric polynomial takes
real values around the unit circle, i.e. $R\left(e^{i\theta}\right)\in\mathbb{R}$
for all $\theta\in\left[0,2\pi\right)$, if and only if vector $r\in\mathbb{C}^{\bar{n}}$
satisfies the Hermitian symmetry condition
\begin{equation}
\forall k\in\left\llbracket 0,n-1\right\rrbracket ,\quad r_{-k}=\overline{r_{k}}.\label{eq:TrigonometricPolynomial-RealValueCondition}
\end{equation}
Every element of $\mathbb{C}^{\bar{n}}\left[X\right]$ can be associated
with a subset of $\mathrm{M}_{n}\left(\mathbb{C}\right)$, called
Gram set, as defined bellow.
\begin{defn}
\label{def:GramSet}A complex matrix $G\in\mathrm{M}_{n}\left(\mathbb{C}\right)$
is a \emph{Gram matrix} associated with the trigonometric polynomial
$R$ if and only if
\[
\forall z\in\mathbb{C}^{*},\quad R\left(z\right)=\psi_{n}\left(z^{-1}\right)^{\trans}G\psi_{n}\left(z\right).
\]
Such parametrization is, in general, not unique and we denote by $\mathcal{G}\left(R\right)$
the set of matrices satisfying the above relation. $\mathcal{G}\left(R\right)$
is called \emph{Gram set} of $R$.
\end{defn}
The next proposition characterizes the Gram set of a complex trigonometric
polynomial taking real values on the unit circle via a simple linear
relation.
\begin{prop}
\label{prop:GramParametrizationTheorem}Let $R\in\mathbb{C}^{\bar{n}}\left[X\right]$
if a complex trigonometric polynomial taking real values around the
unit circle. Let $G\in\mathrm{M}_{n}\left(\mathbb{C}\right)$, then
$G\in\mathcal{G}\left(R\right)$ if and only if the relation
\[
\mathcal{T}_{n}^{*}\left(G\right)=r
\]
holds, where $r=\left[r_{0},\dots,r_{n-1}\right]^{\trans}\in\mathbb{C}^{n}$
is the vector containing the coefficients of $R$ corresponding to
its positive exponents.
\end{prop}
The interested reader is invited to refer to \cite[Theorem 2.3]{Dumitrescu2010}
for a proof and further consequences of this proposition. 

\subsection{Compact representations of polynomials in subspaces}

The notion of Gram sets adapts to every complex trigonometric polynomial;
if $R$ is of order $\bar{n}$, it defines a subset $\mathcal{G}\left(R\right)$
of matrices from $\mathrm{M}_{n}\left(\mathbb{C}\right)$. In our
context, the polynomials of interest have to belong to a low dimensional
subspace characterized by the sub-sampling matrix $M\in\mathbb{C}^{m\times n}$.
Finding compact Gram representations, involving matrices of lower
dimensions, is of crucial interest for reflecting the low dimensionality
of Problem (\ref{eq:ReducedSDP}). In the following, Definition \ref{def:CompactGramRepresentation}
introduces the notion of compact representations, and Corollary \ref{cor:CompactGramParametrizationTheorem}
derives an immediate characterization of those when the considered
polynomial takes real values around the unit circle.
\begin{defn}
\label{def:CompactGramRepresentation}A complex trigonometric polynomial
$R\in\mathbb{C}^{\bar{n}}\left[X\right]$ is said to admit a \emph{compact
Gram representation} on a matrix $M\in\mathbb{C}^{m\times n}$, $m\leq n$
if and only if there exists a matrix $G\in\mathrm{M}_{m}\left(\mathbb{C}\right)$
such that the relation
\begin{align*}
\forall z\in\mathbb{C}^{*},\quad R\left(z\right) & =\psi_{n}\left(z^{-1}\right)^{\trans}M^{\herm}GM\psi_{n}\left(z\right)\\
 & =\phi_{M}\left(z^{-1}\right)^{\trans}G\phi_{M}\left(z\right)
\end{align*}
holds, where $\phi_{M}\left(z\right)=M^{\herm}\psi_{n}\left(z\right)$.
We denote by $\mathcal{G}_{M}\left(R\right)$ the subset of complex
matrices satisfying this property.
\end{defn}
\begin{cor}
\label{cor:CompactGramParametrizationTheorem}Let $R\in\mathbb{C}^{\bar{n}}\left[X\right]$
be a complex trigonometric polynomial taking real values around the
unit circle. Let $G\in\mathrm{M}_{n}\left(\mathbb{C}\right)$, then
$G\in\mathcal{G}_{M}\left(R\right)$ if and only if the relation
\[
\mathcal{R}_{M}^{*}\left(G\right)=r
\]
holds, where $r=\left[r_{0},\dots,r_{n-1}\right]^{\trans}\in\mathbb{C}^{n}$
is the vector containing the coefficients of $R$ corresponding to
its positive exponents.
\end{cor}
The proof of this corollary is a direct consequence of Proposition
\ref{prop:GramParametrizationTheorem} and of the definition of $\mathcal{R}_{M}^{*}$
given in Section \ref{subsec:Notations}.

\subsection{Bounded real lemma for polynomial subspaces\label{subsec:Real-bounded-lemma}}

This part aims to demonstrate a novel result, synthesized in Theorem
\ref{thm:BRLSparsePol}, giving a low-dimensional semidefinite equivalence
of the condition $\left|Q\left(e^{i2\pi\nu}\right)\right|\leq\left|P\left(e^{i2\pi\nu}\right)\right|$
for all $\nu\in\mathbb{T}$ when $P$ and $Q$ are complex polynomials
whose respective coefficients vectors $p,q\in\mathbb{C}^{n}$ lie
in the range of a linear operator $M^{\herm}$, where $M\in\mathbb{C}^{m\times n}$.
Before going into its statement, it is necessary to introduce the
intermediate Proposition \ref{prop:JointSparseGramRep} which highlights
the compatibility between canonical partial order relations defined
on the set trigonometric polynomials and the one of Hermitian matrices.
\begin{prop}
\label{prop:JointSparseGramRep}Let $R\in\mathbb{C}^{\bar{n}}\left[X\right]$
and $R^{\prime}\in\mathbb{C}^{\bar{n}}\left[X\right]$ be two complex
trigonometric polynomials taking real values around the unit circle.
Let by $M\in\mathbb{C}^{m\times n}$ a full rank matrix and suppose
that the sets $\mathcal{G}_{M}\left(R\right)$ and $\mathcal{G}_{M}\left(R^{\prime}\right)$
are both non-empty. Then the inequality $R^{\prime}\left(e^{i2\pi\nu}\right)\leq R\left(e^{i2\pi\nu}\right)$
holds for all $\nu\in\mathbb{T}$ if and only if for every two Hermitian
matrices $G\in\mathcal{G}_{M}\left(R\right)$ and $G^{\prime}\in\mathcal{G}_{M}\left(R^{\prime}\right)$,
one has $G^{\prime}\preceq G$.
\end{prop}
The proof of Proposition \ref{prop:JointSparseGramRep} is provided
in Appendix \ref{sec:GramSet-Polynom}. We are now able to state and
demonstrate a generic algebra result, linking the dominance around
the unit circle of polynomials belonging to some subspace of $\mathbb{C}^{n-1}\left[X\right]$
with an Hermitian semidefinite inequality. Theorem \ref{thm:BRLSparsePol}
plays a key role in the demonstration of Theorem \ref{thm:Dual-CompactSDP-Equivalence}.
\begin{thm}
[Constrained Bounded Real Lemma]\label{thm:BRLSparsePol} Let $P$
and $Q$ be two polynomials of $\mathbb{C}^{n-1}\left[X\right]$ with
respective coefficients vectors $p,q\in\mathbb{C}^{n}$. Moreover,
suppose that $p$ and $q$ belong to the range of $M^{\herm}$, where
$M\in\mathbb{C}^{m\times n}$ is a full rank matrix, and denote by
$u\in\mathbb{C}^{m}$ a vector satisfying $q=M^{\herm}u$. Define
by $R$ the trigonometric polynomial $R\left(z\right)=P\left(z^{-1}\right)P^{*}\left(z\right)$
for all $z\in\mathbb{C}^{*}$, and call $r\in\mathbb{C}^{n}$ its
positive coefficients such that $R$ can be written under the form
$R\left(z\right)=r{}_{0}+\sum_{k=1}^{n-1}\left(r_{k}z^{k}+\overline{r_{k}}z^{-k}\right)$
for all $z\in\mathbb{C}^{*}$. Then the inequality
\[
\forall\nu\in\mathbb{T},\quad\left|Q\left(e^{i2\pi\nu}\right)\right|\leq\left|P\left(e^{i2\pi\nu}\right)\right|
\]
holds if and only if there exists a matrix $S\in\mathrm{S}_{m}\left(\mathbb{C}\right)$
verifying
\[
\begin{cases}
\begin{bmatrix}S & u\\
u^{\herm} & 1
\end{bmatrix}\succeq0\\
\mathcal{R}_{M}^{*}\left(S\right)=r.
\end{cases}
\]
\end{thm}
\begin{IEEEproof}
Denote by $R^{\prime}$ the trigonometric polynomial defined by $R^{\prime}\left(z\right)=Q\left(z^{-1}\right)Q^{\herm}\left(z\right)$
for all $z\in\mathbb{C}^{*}$. Since the identities $R^{\prime}\left(e^{i2\pi\nu}\right)=\left|Q\left(e^{-i2\pi\nu}\right)\right|^{2}$
and $R\left(e^{i2\pi\nu}\right)=\left|P\left(e^{-i2\pi\nu}\right)\right|^{2}$
are verified for all $\nu\in\mathbb{T}$, the inequality $\left|Q\left(e^{i2\pi\nu}\right)\right|\leq\left|P\left(e^{i2\pi\nu}\right)\right|$
is equivalent to $R^{\prime}\left(e^{i2\pi\nu}\right)\leq R\left(e^{i2\pi\nu}\right)$
for all $\nu\in\mathbb{T}$. In the latter, we derive conditions for
this second inequality to hold.

First of all, since $p$ is the range of $M^{\herm}$, one can find
a vector $v\in\mathbb{C}^{m}$ verifying $p=M^{\herm}v$. It comes
that
\begin{align*}
\forall\nu\in\mathbb{T},\quad R\left(e^{i2\pi\nu}\right) & =P\left(e^{-i2\pi\nu}\right)P^{*}\left(e^{i2\pi\nu}\right)\\
 & =\psi_{n}\left(e^{i2\pi\nu}\right)^{*}pp^{*}\psi_{n}\left(e^{i2\pi\nu}\right)\\
 & =\psi_{n}\left(e^{i2\pi\nu}\right)^{*}M^{\herm}vv^{\herm}M\psi_{n}\left(e^{i2\pi\nu}\right).
\end{align*}
Thus, the rank one matrix $vv^{\herm}$ belongs to $\mathcal{G}_{M}\left(R\right)$.
On a similar manner, one has $uu^{\herm}\in\mathcal{G}_{M}\left(R^{\prime}\right)$
and the sets $\mathcal{G}_{M}\left(R\right)$ and $\mathcal{G}_{M}\left(R^{\prime}\right)$
are non-empty. The conditions of Proposition \ref{prop:JointSparseGramRep}
are met. Consequently, the inequality $R^{\prime}\left(e^{i2\pi\nu}\right)\leq R\left(e^{i2\pi\nu}\right)$
holds for all $\nu\in\mathbb{T}$ if and only if there exists a Hermitian
matrix $S\in\mathcal{G}_{\mathcal{I}}\left(R\right)$ satisfying $S\succeq uu^{\herm}$.
By Corollary \ref{cor:CompactGramParametrizationTheorem}, $S\in\mathcal{G}_{\mathcal{I}}\left(R\right)$
is equivalent to $\mathcal{R}_{M}^{*}\left(S\right)=r$. Moreover,
making use of the Schur complement, one has
\[
S\succeq uu^{\herm}\Leftrightarrow\begin{bmatrix}S & u\\
u^{\herm} & 1
\end{bmatrix}\succeq0,
\]
which concludes on the desired result.
\end{IEEEproof}

\subsection{Proof of the main statement}

We conclude in this section by proving that the dual SDP (\ref{eq:FullSDP})
is equivalent to a compact one (\ref{eq:ReducedSDP}) whenever the
sub-sampling operator $M\in\mathbb{C}^{m\times n}$ is admissible
in the sense of Definition \ref{def:AdmissibleOperator}. The proof
of this result is a consequence of the constraint bounded real lemma
presented in the previous Section \ref{subsec:Real-bounded-lemma}. 
\begin{IEEEproof}
By Lemma \ref{lem:DualCharacterization} the dual feasible set $\mathcal{D}_{M}$
of the relaxed problem (\ref{eq:GenericPartialL1}) writes
\[
\mathcal{D}_{M}=\left\{ c\in\mathbb{C}^{m},\;\begin{cases}
q=M^{\herm}c\\
\left\Vert Q\left(e^{i2\pi\nu}\right)\right\Vert _{\infty}\leq1
\end{cases}\right\} ,
\]
where $Q\in\mathbb{C}^{n-1}\left[X\right]$ is the polynomial having
for coefficients vector $q\in\mathbb{C}^{n}$. The core idea of the
proof consist in recasting the inequality on the infinite norm of
$Q$ by
\[
\forall\nu\in\mathbb{T},\quad\left|Q\left(e^{i2\pi\nu}\right)\right|\leq\left|P_{1}\left(e^{i2\pi\nu}\right)\right|,
\]
where $P_{1}$ is the constant unitary polynomial of $\mathbb{C}^{n-1}\left[X\right]$.
Define by $R_{1}$ the constant complex trigonometric polynomial of
$\mathbb{C}^{\bar{n}}\left[X\right]$ reading $R_{1}\left(z\right)=P_{1}\left(z^{-1}\right)P_{1}^{*}\left(z\right)=1$
for all $z\in\mathbb{C}^{*}$. The vector $r_{1}\in\mathbb{C}^{n}$
of its positive monomial coefficients writes $r_{1}=e_{0}$.

For any $c\in\mathcal{D}_{M}$, the vector $q=M^{*}c$ belongs to
the range of $M^{\herm}$. Moreover, since $M$ is admissible, $r_{1}=e_{0}\in\range\left(M^{*}\right)$.
The condition of application of Theorem \ref{thm:BRLSparsePol} are
met, and the equivalence
\[
c\in\mathcal{D}_{M}\Leftrightarrow\exists S\,\text{Hermitian s.t. }\begin{cases}
\begin{bmatrix}S & c\\
c^{\herm} & 1
\end{bmatrix}\succeq0\\
\mathcal{R}_{M}^{*}\left(S\right)=e_{0}
\end{cases}
\]
holds, which concludes the demonstration.
\end{IEEEproof}

\section{Discussion and future work}

The construction of dual polynomial $Q_{\star}$ matching the conditions
(\ref{eq:DualCertificateCondition}) has been successfully achieved
in the partial observation case only for very specific categories
of sub-sampling matrices. It would be of great interest to characterize
more finely the conditions for the existence of such polynomials in
the generic case. In particular, highlighting the loss of resolution
induced by the choice of the sub-sampling matrix $M$ can have an
impact in understanding the trade-off between the heavy high resolution
recovery provided by the full measurement framework, and the fast
coarser estimate provided by partial sub-sampling. However, such approaches
would require to restrict the construction of the dual polynomial
proposed in \cite{Candes2014a} to subspaces formed by non-aligned
observations, which can be technically challenging.

Finally, we suggested in Remark \ref{rem:CommonGridApproximation}
that $\varepsilon$-approximating common grid could be used as an
approximation when the conditions of Proposition \ref{prop:ExistenceOfCommonSupportingGrid}
do not strictly hold, and proposed to consider their performances
under the lens of an analogue basis mismatch problem. Since the dimensionality
of the reduced SDP (\ref{eq:ReducedSDP}) recovering the frequencies
does not depend on the size of the common grid, one can wonder how
the proofs presented in this paper can extend to a super-resolution
theory of sparse spectrum from fully asynchronous measurements by
letting the observation operator $\mathcal{F}_{kn,kf}$ deviating
when $k$ grows large.

\appendices{}

\section{Proof of Proposition \ref{prop:JointSparseGramRep}\label{sec:GramSet-Polynom}}

We start the demonstration by proving the following lemma.
\begin{lem}
\label{lem:EquivalencePositivityWithPositiveHermitian}Let $R\in\mathbb{C}^{\bar{n}}\left[X\right]$
be a complex trigonometric polynomial. Let $M\in\mathbb{C}^{m\times n}$,
$m\leq n$ be a full rank matrix, and suppose that $\mathcal{G}_{M}\left(R\right)$
is not empty. The following assertions hold:

$R$ takes real values around the unit circle if and only if $\mathcal{G}_{M}\left(R\right)$
intersects the set of Hermitian matrices, i.e.
\begin{equation}
\forall\nu\in\mathbb{T},\quad R\left(e^{i2\pi\nu}\right)\in\mathbb{R}\Leftrightarrow\mathcal{G}_{M}\left(R\right)\cap\mathrm{S}_{m}\left(\mathbb{C}\right)\neq\emptyset.\label{eq:Assert-RealToHermitian}
\end{equation}

$R$ takes positive values around on the unit circle if and only if
$\mathcal{G}_{M}\left(R\right)$ intersects the cone of positive Hermitian
matrices, i.e.
\begin{equation}
\forall\nu\in\mathbb{T},\quad R\left(e^{i2\pi\nu}\right)\in\mathbb{R}^{+}\Leftrightarrow\mathcal{G}_{M}\left(R\right)\cap\mathrm{S}_{m}^{+}\left(\mathbb{C}\right)\neq\emptyset,\label{eq:Assert-RealPosToHermitianPos}
\end{equation}
and every Hermitian matrix in $\mathcal{G}_{\mathcal{I}}\left(R\right)$
is positive.
\end{lem}
\begin{IEEEproof}
We start the demonstration by showing that the set $\mathcal{G}_{M}\left(R\right)$
is a convex set. The proof is immediate by taking any two matrices
$G$ and $G^{\prime}$ in $\mathcal{G}_{M}\left(R\right)$ and any
real $\beta\in\left[0,1\right]$. Recalling the definition of the
compact Gram set, it yields
\begin{align*}
\forall z\in\text{\ensuremath{\mathbb{C}}}^{*},\;\phi_{M}\left(z^{-1}\right)^{\trans}\left(\beta G+\left(1-\beta\right)G^{\prime}\right)\phi_{M}\left(z\right) & =\beta R\left(z\right)+\left(1-\beta\right)R\left(z\right)\\
 & =R\left(z\right).
\end{align*}
Thus $\beta G+\left(1-\beta\right)G^{\prime}\in\mathcal{G}_{\mathcal{I}}\left(R\right)$,
and the convexity follows.

We carry on the demonstration of Assertion (\ref{eq:Assert-RealToHermitian})
by showing that $R$ takes real values around the unit circle if and
only if the set $\mathcal{G}_{\mathcal{I}}\left(R\right)$ is stable
by Hermitian transposition. First of all, it is easy to see via Definition
\ref{def:GramSet} of the set $\mathcal{\mathcal{G}}_{M}\left(R\right)$
that 
\[
G\in\mathcal{\mathcal{G}}_{M}\left(R\right)\Leftrightarrow G^{\herm}\in\mathcal{G}_{M}\left(R^{*}\right).
\]
Moreover since $R$ takes real values around the unit circle, its
coefficient vector satisfies the symmetry property (\ref{eq:TrigonometricPolynomial-RealValueCondition})
(and reciprocally), which translate into $\mathcal{\mathcal{G}}_{M}\left(R\right)=\mathcal{\mathcal{G}}_{M}\left(R^{\herm}\right)$.
Combining the last two relations lead to the equivalence with the
stability of $\mathcal{G}_{M}\left(R\right)$ by Hermitian transposition,
i.e.
\begin{equation}
R\left(e^{i2\pi\nu}\right)\in\mathbb{R}\Leftrightarrow\forall G\in\mathcal{\mathcal{G}}_{M}\left(R\right),\;G^{\herm}\in\mathcal{\mathcal{G}}_{M}\left(R\right).\label{eq:Assert-StabilityByHermitianTranspose}
\end{equation}
We conclude the demonstration of Assertion (\ref{eq:Assert-RealToHermitian})
by taking any element $G\in\mathcal{\mathcal{G}}_{M}\left(R\right)$,
and by noticing
\[
R\left(e^{i2\pi\nu}\right)\in\mathbb{R}\Leftrightarrow\forall G\in\mathcal{\mathcal{G}}_{M}\left(R\right),\;\frac{G+G^{\herm}}{2}\in\mathcal{\mathcal{G}}_{M}\left(R\right),
\]
using the convexity and the stability of $\mathcal{\mathcal{G}}_{M}\left(R\right)$
by Hermitian transposition. Since $\mathcal{\mathcal{G}}_{M}\left(R\right)$
is not empty by assumption, it intersects non-trivially the set of
Hermitian matrices.

Suppose now that $R$ takes real positive values over the unit circle.
Let by $S$ a Hermitian matrix belonging to $\mathcal{G}_{M}\left(R\right)$
(Assertion (\ref{eq:Assert-RealToHermitian}) attests the existence
of such matrix). It comes
\begin{align*}
\forall\nu\in\mathbb{T},\quad R\left(e^{i2\pi\nu}\right) & =\psi_{n}\left(e^{-i2\pi\nu}\right)^{\trans}M^{\herm}SM\psi_{n}\left(e^{i2\pi\nu}\right)\\
 & =\phi_{M}\left(e^{-i2\pi\nu}\right)^{\herm}S\phi_{M}\left(e^{i2\pi\nu}\right).
\end{align*}
Since the sub-sampling matrix $M$ is full rank, the set $\left\{ \phi_{M}\left(e^{i2\pi\nu}\right),\;\nu\in\mathbb{T}\right\} $
spans the whole vectorial space $\mathbb{C}^{m}$. Thus, the positivity
of $R$ is equivalent to the positivity of the Hermitian matrix $S$,
concluding on the second statement of the lemma.
\end{IEEEproof}
We are now ready to start the demonstration the Proposition \ref{prop:JointSparseGramRep}.

Denote respectively by $r,r^{\prime}\in\mathbb{C}^{n}$ the respective
positive coefficients of the trigonometric polynomials $R$ and $R^{\prime}$.
The sets $\mathcal{G_{I}}\left(R\right)$ and $\mathcal{G_{I}}\left(R^{\prime}\right)$
are non-empty by assumption, and Lemma \ref{lem:EquivalencePositivityWithPositiveHermitian}
guarantees the existence of two Hermitian matrices $S_{0}$ and $S_{0}^{\prime}$
belonging respectively to $\mathcal{G_{I}}\left(R\right)$ and $\mathcal{G_{I}}\left(R^{\prime}\right)$.
Define by $T$ the trigonometric polynomial
\begin{align}
\forall\nu\in\mathbb{T},\quad T\left(e^{i2\pi\nu}\right) & =R\left(e^{i2\pi\nu}\right)-R^{\prime}\left(e^{i2\pi\nu}\right)\label{eq:DefOfT}\\
 & =\phi_{M}\left(e^{i2\pi\nu}\right)^{\herm}\left(S_{0}-S_{0}^{\prime}\right)\phi_{M}\left(e^{i2\pi\nu}\right).\nonumber 
\end{align}
Proving that $R$ is greater than $R^{\prime}$ around the unit circle
is equivalent to prove the positivity of $T$ on the same domain.
It is clear that the matrix $S_{0}-S_{0}^{\prime}$ belongs to $\mathcal{G}_{M}\left(T\right)$
and thus $\mathcal{G}_{M}\left(T\right)$ is not empty. By application
of Lemma \ref{lem:EquivalencePositivityWithPositiveHermitian}, $T$
is positive if and only if every Hermitian matrix $H$ in the set
$\mathcal{G}_{M}\left(T\right)$ is positive. We conclude that $T$
is positive if and only if for every pair of Hermitian matrices $\left(S,S^{\prime}\right)\in\mathcal{G}_{\mathcal{I}}\left(R\right)\times\mathcal{G}\left(R'\right)$
one has $S\succeq S^{\prime}$.\hfill{}$\blacksquare$

\section{Dual characterization lemma \ref{lem:DualCharacterization} and proof
of Proposition \ref{prop:PolynomialAlignment}}

\subsection{Proof of Lemma \ref{lem:DualCharacterization} }

A standard Lagrangian analysis leads to a dual of $\eqref{eq:GenericPartialL1}$
of the form
\begin{align}
c_{\star} & =\arg\max_{c\in\mathbb{C}^{m}}\Re\left(y^{\trans}c\right)\label{eq:RawDualForm}\\
\textrm{subject to} & \phantom{\phantom{\;=\;}}\left\Vert \mathcal{\mathcal{F}}_{n}^{\herm}\left(M^{\herm}c\right)\right\Vert _{\infty}\leq1\nonumber \\
 & \phantom{\;=\;}q=M^{\herm}c.\nonumber 
\end{align}
By direct calculation, one has
\begin{align*}
\forall c\in\mathbb{C}^{m},\forall\xi\in\mathbb{R},\quad\mathcal{F}_{n}^{*}\left(M^{\herm}c\right)\left(\xi\right) & =\mathcal{F}_{n}^{*}\left(q\right)\left(\xi\right)\\
 & =\sum_{k\in\mathcal{I}}q_{k}e^{-i2\pi k\frac{\xi}{f}}\\
 & =Q\left(e^{-i2\pi\frac{\xi}{f}}\right)
\end{align*}
where $q=M^{\herm}c$ is the coefficients vector of the polynomial
$Q\in\mathbb{C}^{n-1}\left[X\right]$. The characterization of $\mathcal{D}_{M}$
follows by noticing the invariance of the infinite norm over the transform
$\xi\leftarrow-\xi$. The equivalence between Program (\ref{eq:RawDualForm})
and an SDP is a direct consequence of the relation
\[
\left\Vert Q\left(e^{i2\pi\nu}\right)\right\Vert _{\infty}\leq1\Leftrightarrow\exists H\text{ Hermitian s.t. }\begin{cases}
\begin{bmatrix}H & q\\
q^{\herm} & 1
\end{bmatrix}\succeq0\\
\mathcal{T}_{n}^{\herm}\left(H\right)=e_{0}.
\end{cases}
\]
A proof of this last assersion can be found in \cite[Corollary 4.25]{Dumitrescu2010}.\hfill{}$\blacksquare$

\subsection{Proof of Proposition \ref{prop:PolynomialAlignment}\label{subsec:Proof-of-PolynomialAlignment}}
\begin{IEEEproof}
We recall from Equation (\ref{eq:Def-ObservationOperator}) that for
all $\hat{x}\in D_{1}$, one has,
\[
\forall j\in\left\llbracket 1,m\right\rrbracket ,\forall k\in\left\llbracket 0,n_{j}-1\right\rrbracket ,\quad\linmeasure_{j}\left[k\right]=\int_{\mathbb{R}}e^{i2\pi\frac{\xi}{f_{j}}\left(k-\gamma_{j}\right)}\mathrm{d}\hat{x}\left(\xi\right).
\]

Suppose that $\mathcal{C}\left(\mathbb{A}\right)$ is not empty, the
minimal common supporting grid $\mathcal{A}_{\baro}=\left(f_{\baro},\gamma_{\baro},n_{\baro}\right)$
for $\mathbb{A}$ exists. It comes by Equation (\ref{eq:CommonSupportingGrid-Definition})
that
\begin{align*}
\forall j\in\left\llbracket 1,m\right\rrbracket ,\forall k_{j}\in\left\llbracket 0,n_{j}-1\right\rrbracket ,\exists k_{\baro}\in\left\llbracket 0,n_{\baro}-1\right\rrbracket ,\quad\linmeasure_{j}\left(\hat{x}\right)\left[k_{j}\right] & =\int_{\mathbb{R}}e^{i2\pi\frac{\xi}{f_{\baro}}\left(k_{\baro}-\gamma_{\baro}\right)}\mathrm{d}\hat{x}\left(\xi\right)\\
 & =\int_{\mathbb{R}}e^{i2\pi\frac{\xi}{f_{\baro}}k_{\baro}}\mathrm{d}\left(e^{-i2\pi\frac{\xi\gamma_{\baro}}{f_{\baro}}}\hat{x}\left(\xi\right)\right)\\
 & =\mathcal{F}_{n_{\baro}}\circ\mathcal{M}_{\frac{\gamma_{\baro}}{f_{\baro}}}\left(\hat{x}\right)\left[k\right].
\end{align*}
Let by $\mathcal{I}\subseteq\left\llbracket 0,n_{\baro}-1\right\rrbracket $
the equivalent observation set of the minimal $\mathcal{A}_{\baro}$
introduced in Definition \ref{def:CommonSupportingGrid} and consider
a selection matrix $C_{\mathcal{I}}\in\mathbb{C}^{m\times n_{\baro}}$
for this set. The above equality ensures the measurement operator
admit a factorization of the form 
\[
\linmeasure=C_{\mathcal{I}}\left(\mathcal{F}_{n_{\baro},f_{\baro}}\circ\mathcal{M}_{\frac{\gamma_{\baro}}{f_{\baro}}}\right).
\]
Finally, $0\in\mathcal{I}$ by minimality of the grid $\mathcal{A}_{\baro}$,
and the selection matrix $C_{\mathcal{I}}$ is an admissible sub-sampling
operator in the sense of Definition \ref{def:AdmissibleOperator}.

Since any selection matrix $C_{\mathcal{I}}\in\mathbb{C}^{m\times n}$
can be interpreted as a MRSS with $m$ aligned grids taking a single
sample ($n_{j}=1$ for all $j\in\left\llbracket 1,m\right\rrbracket $),
the proof of the converse is immediate.
\end{IEEEproof}

\section{Proof of Theorem \ref{thm:MRSSDualCertifiability}\label{subsec:Proof-of-DualCertifiability}}

In both strong and weak condition cases, the proof relies on previous
works presented in \cite{Candes2014a,Tang2013}, and is achieved by
constructing a polynomial $Q_{\star}$ satisfying the conditions (\ref{eq:DualCertificateCondition}).
It is been shown in Section \ref{subsec:Common-grid-expansion} that
shifting the signal in the time domain leave the dual feasible set
invariant, and we will assume without loss of generality that $\gamma_{\baro}=0$
so that $\linmeasure=C_{\mathcal{I}}\mathcal{F}_{n}$. Before starting
the proof, we introduce the notations
\begin{align*}
\Omega_{\baro} & =\frac{1}{f_{\baro}}\Xi=\left\{ \frac{\xi}{f_{\baro}},\;\xi\in\Xi\right\} \\
\forall j\in\left\llbracket 1,p\right\rrbracket ,\quad\Omega_{j} & =\frac{1}{f_{j}}\Xi=\left\{ \frac{\xi}{f_{j}},\;\xi\in\Xi\right\} \\
\forall j\in\left\llbracket 1,p\right\rrbracket ,\quad\tilde{\Omega}_{j} & =\left\{ \frac{\xi}{f_{\baro}}+\frac{k}{l_{j}},\;\xi\in\Xi,k\in\left\llbracket 0,l_{j}-1\right\rrbracket \right\} .
\end{align*}
In the above, $\Omega_{\baro}$ and $\Omega_{j}$ are the sets of
the reduced frequencies of the spectral support $\Xi$ of the signal
$x$ for the respective sampling frequencies $f_{\baro}$ and $f_{j}$,
while $\tilde{\Omega}_{j}$ is the aliased set of $\Omega_{j}$ resulting
from a zero-forcing upsampling from the rate $f_{j}$ to the rate
$f_{\baro}$.

We recall from \cite{Tang2013} Proposition II.4, using the improved
separability conditions taken from \cite{Fernandez-granda2015} Proposition
4.1, that if $\Delta_{\mathbb{T}}\left(\Omega_{j}\right)\geq\frac{2.52}{n_{j}-1}$,
then one can build a polynomial $P_{j,\star}\in\mathbb{C}^{n_{j}-1}\left[X\right]$
satisfying the interpolating conditions
\begin{equation}
\begin{cases}
P_{j,\star}\left(e^{i2\pi\frac{\xi_{r}}{f_{j}}}\right)=\sign\left(e^{i2\pi\frac{a_{j}}{l_{j}}\frac{\xi_{r}}{f_{j}}}\alpha_{r}\right), & \forall\frac{\xi_{r}}{f_{j}}\in\Omega_{j}\\
\left|P_{j,\star}\left(e^{i2\pi\nu}\right)\right|<1, & \forall\nu\in\mathbb{T}\backslash\Omega_{j}\\
\frac{\mathrm{d}^{2}\left|P_{j,\star}\right|}{\mathrm{d}\nu^{2}}\left(e^{i2\pi\frac{\xi_{r}}{f_{j}}}\right)\leq-\eta, & \forall\frac{\xi_{r}}{f_{j}}\in\Omega_{j},
\end{cases}\label{eq:individualCertificate}
\end{equation}
provided that $n_{j}>2\times10^{3}$, for some $\eta>0$ ($\eta=7.865\,10^{-2}$
in the original proof presented in \cite{Fernandez-granda2015}),
and whereby $\left\{ \left(a_{j},l_{j}\right)\right\} _{j\in\left\llbracket 1,p\right\rrbracket }$
are the pairs of parameters defined in the statement of Proposition
\ref{prop:ExistenceOfCommonSupportingGrid} characterizing the expansion
of the array $\mathcal{A}_{j}$ into the minimal common grid $\mathcal{A}_{\baro}$.
If the polynomial $P_{j,\star}$ exists, we further introduce the
polynomial $Q_{j,\star}\in\mathbb{C}^{n_{\baro}-1}\left[X\right]$
defined by
\begin{equation}
\forall z\in\mathbb{C},\quad Q_{j,\star}\left(z\right)=z^{-a_{j}}P_{j,\star}\left(z^{l_{j}}\right).\label{eq:polynomialUpscaling}
\end{equation}
By construction, $Q_{j,\star}$ is a sparse polynomial with monomial
support on the subset $\mathcal{I}$ introduced in Proposition \ref{prop:PolynomialAlignment}.
Its coefficients vector $q_{j,\star}$ satisfies the relation $q_{j,\star}=C_{\mathcal{I}}^{\herm}c_{j,\star}$
for some $c_{j,\star}\in\mathbb{C}^{m}$. It is easy to notice that
due to the upscaling effect $z\leftarrow z^{l_{j}}$ in (\ref{eq:polynomialUpscaling})
the function
\begin{align*}
\mathbb{R} & \rightarrow\mathbb{C}\\
\nu & \mapsto\left|Q_{j,\star}\left(e^{i2\pi\nu}\right)\right|,
\end{align*}
is $\frac{1}{l_{j}}$-periodic. Consequently the polynomial $Q_{j,\star}$
reaches a modulus equal to $1$ on every point of $\tilde{\Omega}_{j}$,
with value satisfying

\begin{align*}
\forall\nu\in\tilde{\Omega}_{j},\quad Q_{j}\left(e^{i2\pi\nu}\right) & =Q_{j,\star}\left(e^{i2\pi\left(\frac{\xi_{r}}{f_{\baro}}+\frac{k}{l_{j}}\right)}\right)\\
 & =e^{-i2\pi a_{j}\left(\frac{\xi_{r}}{f_{\baro}}+\frac{k}{l_{j}}\right)}P_{j,\star}\left(e^{i2\pi\left(\frac{l_{j}\xi_{r}}{f_{\baro}}+k\right)}\right)\\
 & =e^{-i2\pi a_{j}\left(\frac{\xi_{r}}{f_{\baro}}+\frac{k}{l_{j}}\right)}\sign\left(e^{i2\pi\frac{a_{j}}{l_{j}}\frac{\xi_{r}}{f_{j}}}\alpha_{r}\right)\\
 & =e^{-i2\pi a_{j}\frac{k}{l_{j}}}\sign\left(\alpha_{r}\right),
\end{align*}
whereby $\frac{\xi_{r}}{f_{\baro}}\in\Omega_{\baro}$ and $k\in\left\llbracket 0,l_{j}-1\right\rrbracket $.
It comes that the constructed polynomial verifies the interpolation
conditions

\begin{equation}
\begin{cases}
Q_{j,\star}\left(e^{i2\pi\nu}\right)=\sign\left(\alpha_{r}\right), & \forall\nu\in\Omega_{\baro}\\
Q_{j,\star}\left(e^{i2\pi\nu}\right)=e^{-i2\pi a_{j}\frac{k}{l_{j}}}\sign\left(\alpha_{r}\right), & \forall\nu\in\tilde{\Omega}_{j}\\
\left|Q_{j,\star}\left(e^{i2\pi\nu}\right)\right|<1, & \forall\nu\in\mathbb{T}\backslash\tilde{\Omega}_{j}\\
\frac{\mathrm{d}^{2}\left|Q_{j,\star}\right|}{\mathrm{d}\nu^{2}}\left(e^{i2\pi\nu}\right)\leq-l_{j}\eta, & \forall\nu\in\tilde{\Omega}_{j},
\end{cases}\label{eq:upscaledInterpolationProperties}
\end{equation}
where the second equality stand for some $\frac{\xi_{r}}{f_{\baro}}\in\Omega_{\baro}$
and $k\in\left\llbracket 0,l_{j}-1\right\rrbracket $ such that $\nu=\frac{\xi_{r}}{f_{\baro}}+\frac{k}{l_{j}}\in\tilde{\Omega}_{j}$.

Under both strong and weak assumptions, we aim to build a sparse polynomial
$Q_{\star}\in\mathbb{C}^{n_{\baro}-1}\left[X\right]$ verifying the
conditions (\ref{eq:DualCertificateCondition}). If the existence
of such polynomial is verified \ref{prop:DualCertifiability} applies
and the desired conclusion follows.
\begin{IEEEproof}
[Construction under the strong condition]Suppose that $\Delta_{\mathbb{T}}\left(\Omega_{j}\right)\geq\frac{2.52}{n_{j}-1}$
and $n_{j}>2\times10^{3}$, for all $j\in\left\llbracket 1,p\right\rrbracket $,
as explained above, one can find $p$ polynomials $Q_{j,\star}\in\mathbb{C}^{n_{\baro}-1}\left[X\right]$
satisfying the interpolation properties given in (\ref{eq:upscaledInterpolationProperties}).
Define by $Q_{\star}\in\mathbb{C}^{n_{\baro}-1}\left[X\right]$ their
average
\[
\forall z\in\mathbb{C},\quad Q_{\star}\left(z\right)=\frac{1}{p}\sum_{j=1}^{p}Q_{j,\star}\left(z\right).
\]
It is clear, by stability through linear combinations, that $Q_{\star}$
is still sparse and supported over the subset $\mathcal{I}$, ensuring
the existence of an element $c_{\star}\in\mathbb{C}^{m}$ such that
$q_{\star}=C_{\mathcal{I}}^{\herm}c_{\star}$. Moreover, it is immediate
to verify that $Q_{\star}$ satisfies
\begin{align}
\left|Q_{\star}\left(e^{i2\pi\nu}\right)\right|=1 & \Leftrightarrow\left(\nu\in\bigcap_{j=1}^{p}\tilde{\Omega}_{j}\text{ and }\forall j\in\left\llbracket 1,p\right\rrbracket ,\;Q_{j,\star}\left(e^{i2\pi\nu}\right)=u\left(\nu\right)\right)\label{eq:Gamma-Definition}
\end{align}
for some value $u\left(\nu\right)\in\mathbb{C}$ of modulus $1$,
$\left|u\left(\nu\right)\right|=1$. Let us denote by $\Gamma\subset\mathbb{T}$
the set of frequencies satisfying (\ref{eq:Gamma-Definition}). From
(\ref{eq:DualCertificateCondition}) and (\ref{eq:upscaledInterpolationProperties}),
$Q_{\star}$ is a dual certificate if and only if $\Gamma=\Omega_{\baro}$.
One has $\Omega_{\baro}\subseteq\Gamma$, thus it remains to prove
$\Gamma\subseteq\Omega_{\baro}$ to finish the certificate construction
under the strong condition. Using the definition of $\tilde{\Omega}_{j}$
and the interpolation properties (\ref{eq:upscaledInterpolationProperties}),
we have that $\nu\in\Gamma$ is equivalent to

\begin{multline*}
\nu\in\bigcap_{j=1}^{p}\tilde{\Omega}_{j}\Longleftrightarrow\forall\left(j,j^{\prime}\right)\in\left\llbracket 1,p\right\rrbracket ^{2},\exists\left(r,r^{\prime}\right)\in\left\llbracket 1,s\right\rrbracket ^{2},\exists k_{j}\in\left\llbracket 0,l_{j}-1\right\rrbracket ,\exists k_{j}\in\left\llbracket 0,l_{j^{\prime}}-1\right\rrbracket ,\\
e^{-i2\pi a_{j}\frac{k_{j}}{l_{j}}}\sign\left(\alpha_{r}\right)=e^{-i2\pi a_{j^{\prime}}\frac{k_{j^{\prime}}}{l_{j^{\prime}}}}\sign\left(\alpha_{r^{\prime}}\right),
\end{multline*}
leading to
\begin{multline}
\nu\in\bigcap_{j=1}^{m}\tilde{\Omega}_{j}\Longleftrightarrow\forall\left(j,j^{\prime}\right)\in\left\llbracket 1,p\right\rrbracket ^{2},\exists\left(r,r^{\prime}\right)\in\left\llbracket 1,s\right\rrbracket ^{2},\exists k_{j}\in\left\llbracket 0,l_{j}-1\right\rrbracket ,\exists k_{j}\in\left\llbracket 0,l_{j^{\prime}}-1\right\rrbracket ,\exists b\in\mathbb{Z},\\
a_{j}\frac{k_{j}}{l_{j}}+\frac{\arg\left(\alpha_{r}\right)}{2\pi}=a_{j^{\prime}}\frac{k_{j^{\prime}}}{l_{j^{\prime}}}+\frac{\arg\left(\alpha_{r^{\prime}}\right)}{2\pi}+b.\label{eq:AliasingSetsIntersectionCondition}
\end{multline}
The equality in the RHS of (\ref{eq:AliasingSetsIntersectionCondition})
may occur for all pairs $\left(j,j^{\prime}\right)\in\left\llbracket 1,p\right\rrbracket ^{2}$
if and only if $r=r^{\prime}$, and the above reduces to
\begin{multline*}
\nu\in\bigcap_{j=1}^{m}\tilde{\Omega}_{j}\Longleftrightarrow\forall\left(j,j^{\prime}\right)\in\left\llbracket 1,p\right\rrbracket ^{2},\exists k\in\left\llbracket 0,l_{j}-1\right\rrbracket ,\exists k^{\prime}\in\left\llbracket 0,l_{j^{\prime}}-1\right\rrbracket ,\exists b\in\mathbb{Z},\quad\frac{a_{j}k_{j}}{l_{j}}=\frac{a_{j^{\prime}}k_{j^{\prime}}}{l_{j^{\prime}}}+b,
\end{multline*}
which holds if and only if
\[
\forall\left(j,j^{\prime}\right)\in\left\llbracket 1,p\right\rrbracket ^{2},\exists k\in\left\llbracket 0,l_{j}-1\right\rrbracket ,\;l_{j}\mid a_{j}l_{j^{\prime}}k_{j}.
\]
\sloppy Recalling from the minimality condition of the common grid
$\mathcal{A}_{\baro}$ detailed in Proposition \ref{prop:ExistenceOfCommonSupportingGrid}
that $\gcd\left(\left\{ a_{j}\right\} _{j\in\left\llbracket 1,p\right\rrbracket }\cup\left\{ l_{j}\right\} _{j\in\left\llbracket 1,p\right\rrbracket }\right)=1$,
one derives by application of the Gauss theorem
\[
\exists j\in\left\llbracket 1,p\right\rrbracket ,\quad l_{j}\mid k_{j}.
\]
\fussy Since $k_{j}\in\left\llbracket 0,l_{j}-1\right\rrbracket $,
one has $k_{j}=0$. We deduce that there must exists $r\in\left\llbracket 1,s\right\rrbracket $
such that $\nu=\frac{\xi_{r}}{f_{\baro}}+\frac{0}{l_{j}}$ and finally
$\nu\in\Omega_{\baro}$. Consequently, $\Gamma\subseteq\Omega_{\baro}$,
and finally $\Gamma=\Omega_{\baro}$, which concludes the proof for
the strong condition.
\end{IEEEproof}
\begin{IEEEproof}
[Construction under the weak condition]Suppose that $\Delta_{\mathbb{T}}\left(\Omega_{j}\right)\geq\frac{2.52}{n_{j}-1}$
and $n_{j}>2\times10^{3}$ for some $j\in\left\llbracket 1,p\right\rrbracket $,
and define the polynomial $Q_{j,\star}\in\mathbb{C}^{n_{\baro}-1}\left[X\right]$
as in Equation (\ref{eq:polynomialUpscaling}). Moreover, we define
by $\mathcal{H}_{j}\left(\mathbb{A},\Omega_{\baro}\right)$ the affine
subspace of elements $c$$\in$$\mathbb{C}^{m}$ such that $q=C_{\mathcal{I}}^{\herm}c$
induces a sparse polynomial $Q\in\mathbb{C}^{n_{\baro}-1}\left[X\right]$
supported by monomials taken over the subset $\mathcal{I}$ and satisfying
the interpolation conditions
\[
\begin{cases}
Q\left(e^{i2\pi\nu}\right)=\sign\left(\alpha_{r}\right), & \forall\nu\in\Omega_{\baro}\\
Q^{\prime}\left(e^{i2\pi\nu}\right)=0, & \forall\nu\in\Omega_{\baro}\\
Q\left(e^{i2\pi\nu}\right)=0, & \forall\nu\in\tilde{\Omega}_{j}\backslash\Omega_{\baro}.
\end{cases}
\]
The subspace $\mathcal{H}_{j}\left(\mathbb{A},\xi\right)$ can be
parametrized by the linear equality
\[
\mathcal{H}_{j}\left(\mathbb{A},\xi\right)=\left\{ c\in\mathbb{C}^{m},V_{j}\left(\mathbb{A},\Omega_{\baro}\right)C_{\mathcal{I}}^{\herm}c=w\right\} ,
\]
whereby $w=\left[\sign\left(\alpha_{1}\right),\dots,\sign\left(\alpha_{s}\right)\right]^{\trans}\in\mathbb{C}^{s}$,
and for some matrix $V_{j}\left(\mathbb{A},\Omega_{\baro}\right)\in\mathbb{C}^{\left(l_{j}+1\right)s\times n_{\baro}}$
defining the interpolation conditions. Interpolation theory guarantees
that $V_{j}\left(\mathbb{A},\Omega_{\baro}\right)$ is full rank,
and therefore the subspace $\mathcal{H}_{j}\left(\mathbb{A},\Omega_{\baro}\right)$
is non-trivial with dimension $m-\left(l_{j}+1\right)s$, provided
that $m\geq\left(l_{j}+1\right)s$. We fix an element $t\in\mathcal{H}_{j}\left(\mathbb{A},\Omega_{\baro}\right)$,
and denote by $R\in\mathbb{C}^{n_{\baro}-1}\left[X\right]$ the polynomial
having for coefficients vector $r=C_{\mathcal{I}}^{\herm}t$. In the
rest of this proof, we seek to build a dual certificate $Q_{\star}\in\mathbb{C}^{n_{\baro}-1}\left[X\right]$
under the form of a convex combination between $R$ and $Q_{j,\star}$
\[
Q_{\star}=\beta R+\left(1-\beta\right)Q_{j,\star},\quad\beta\in\left[0,1\right].
\]

First of all, by construction, $R$ and $Q_{j,\star}$ both interpolate
the frequencies of $\Omega_{\baro}$ with values $w_{r}=\sign\left(a_{r}\right)$,
and one has
\begin{equation}
\forall\nu\in\Omega_{\baro},\quad Q_{\star}\left(e^{i2\pi\nu}\right)=w_{r}.\label{eq:TrueFreqInterpolation}
\end{equation}
Consequently, it remains to derive sufficient conditions on $\beta$
for the optimality condition $\left|Q_{\star}\left(e^{i2\pi\nu}\right)\right|<1$
to hold everywhere else on $\mathbb{T}\backslash\Omega_{\baro}$ to
ensure that $Q_{\star}$ is a dual certificate. To do so, we partition
the set $\mathbb{T}$ into three non-intersecting sets $\mathbb{T}=\Gamma_{\textrm{near}}\cup\Gamma_{\textrm{alias}}\cup\Gamma_{\textrm{far}}$,
where $\Gamma_{\textrm{near}}$ is a union of $s$ open ball of small
radii $0<\varepsilon_{\textrm{near}}$ centered around the frequencies
in $\Omega_{\baro}$, $\Gamma_{\textrm{alias}}$ is an open set containing
the elements of $\tilde{\Omega}_{j}\backslash\Omega_{\baro}$. The
set $\Gamma_{\textrm{far}}$ is defined by the complementary of the
two previous in $\mathbb{T}$. The conditions on $\beta$ for $Q_{\star}$
to be bounded away from $1$ in modulus are derived independently
on each of those sets.

We start the analysis on $\Gamma_{\textrm{near}}$. For any complex
polynomial $Q$, we respectively denote by $Q_{\Re}\left(\nu\right)=\Re\left(Q\left(e^{i2\pi\nu}\right)\right)$
and $Q_{\Im}\left(\nu\right)=\Im\left(Q\left(e^{i2\pi\nu}\right)\right)$
for all $\nu\in\mathbb{T}$, its real and imaginary part around the
unit circle. Moreover, we recall that
\begin{equation}
\frac{\mathrm{d}^{2}\left|Q\right|}{\mathrm{d}\nu^{2}}\left(\nu\right)=-\frac{\left(Q_{\Re}\left(\nu\right)Q_{\Re}^{\prime}\left(\nu\right)+Q_{\Im}\left(\nu\right)Q_{\Im}^{\prime}\left(\nu\right)\right)^{2}}{\left|Q\left(\nu\right)\right|^{3}}+\frac{\left|Q^{\prime}\left(\nu\right)\right|^{2}+Q_{\Re}\left(\nu\right)Q_{\Re}^{\prime\prime}\left(\nu\right)+Q_{\Im}\left(\nu\right)Q_{\Im}^{\prime\prime}\left(\nu\right)}{\left|Q\left(\nu\right)\right|},\label{eq:ModulusSecondDerivative}
\end{equation}
for all $\nu\in\mathbb{T}$. By construction, the derivative of $R$
and $Q_{j,\star}$ cancels on $\Omega_{\baro}$ and by linearity
\begin{equation}
\forall\nu\in\Omega_{\baro},\quad Q_{\star}^{\prime}\left(e^{i2\pi\nu}\right)=0.\label{eq:TrueDerivativeCancelation}
\end{equation}
Injecting Equations (\ref{eq:TrueFreqInterpolation}) and (\ref{eq:TrueDerivativeCancelation})
into (\ref{eq:ModulusSecondDerivative}) leads to
\[
\forall\nu\in\Omega_{\baro},\quad\frac{\mathrm{d}^{2}\left|Q_{\star}\right|}{\mathrm{d}\nu^{2}}\left(\nu\right)=\cos\left(w_{r}\right)Q_{*\Re}^{\prime\prime}\left(\nu\right)+\sin\left(w_{r}\right)Q_{*\Im}^{\prime\prime}\left(\nu\right).
\]
Thus, the operator $\frac{\mathrm{d}^{2}\left|\cdot\right|}{\mathrm{d}\nu^{2}}$
acts linearly on the polynomial $Q_{\star}$ at the points in $\Omega_{\baro}$,
and one has
\begin{align*}
\forall\nu\in\Omega_{\baro},\quad\frac{\mathrm{d}^{2}\left|Q_{\star}\right|}{\mathrm{d}\nu^{2}}\left(\nu\right) & =\beta\frac{\mathrm{d}^{2}\left|R\right|}{\mathrm{d}\nu^{2}}\left(\nu\right)+\left(1-\beta\right)\frac{\mathrm{d}^{2}\left|Q_{j,\star}\right|}{\mathrm{d}\nu^{2}}\left(\nu\right)\\
 & \leq\beta\frac{{\rm d^{2}}\left|R\right|}{{\rm d}\nu^{2}}\left(\nu\right)-\left(1-\beta\right)l_{j}\eta,
\end{align*}
using the interpolation properties of Equation (\ref{eq:upscaledInterpolationProperties}).
The inequalities 
\[
\forall\nu\in\Omega_{\baro},\quad\frac{\mathrm{d}^{2}\left|Q_{\star}\right|}{\mathrm{d}\nu^{2}}\left(\nu\right)<0
\]
can be jointly satisfied, for a choice of $\beta$
\begin{equation}
\beta<\frac{l_{j}\eta}{\mathcal{M}_{\baro}^{\prime\prime}\left(R\right)+l_{j}\eta},\label{eq:betaNearTrue}
\end{equation}
where
\[
\mathcal{M}_{\baro}^{\prime\prime}\left(R\right)=\max_{\nu\in\Omega_{\baro}}\frac{\mathrm{d}^{2}\left|R\right|}{\mathrm{d}\nu^{2}}\left(\nu\right).
\]
Under Condition (\ref{eq:betaNearTrue}) , $\left|Q_{\star}\right|-1$
has $s$ non-nodal roots on $\Omega_{\baro}$, and by continuity of
$Q_{\star}$ there must exist a radius $0<\varepsilon_{\textrm{near}}$
such that
\[
\forall\nu\in\Gamma_{\textrm{near}}\backslash\Omega_{\baro},\quad\left|Q_{\star}\left(e^{i2\pi\nu}\right)\right|<1,
\]
holds where $\Gamma_{\textrm{near}}=\bigcup_{r=1}^{s}\mathcal{B}\left(\frac{\xi_{r}}{f_{\baro}},\varepsilon_{\textrm{near}}\right)$,
where $\mathcal{B}\left(\nu,\varepsilon\right)$ denotes the open
ball of $\mathbb{T}$ of center $\nu$ and radius $\varepsilon$ for
the torus distance.

We continue the proof by bounding $\left|Q_{\star}\right|$ away from
$1$ on the set $\Gamma_{\textrm{alias}}$. Fix any $0<\delta<1$
and let $\Gamma_{\textrm{alias}}=\left\{ \nu,\;\left|R\left(e^{i2\pi\nu}\right)\right|<\delta\right\} $.
By continuity of $R$, $\Gamma_{\textrm{alias}}$ is an open set verifying
$\left(\tilde{\Omega}_{j}\backslash\Omega_{\baro}\right)\subset\Gamma_{\textrm{alias}}$,
moreover one can impose $\Gamma_{\textrm{alias}}\cap\Gamma_{\textrm{near}}=\emptyset$
for a small enough $\delta$. The value of $\left|Q_{\star}\right|$
over $\Gamma_{\textrm{alias}}$ can be bounded by
\begin{align*}
\forall\nu\in\Gamma_{\textrm{alias}},\quad\left|Q_{\star}\left(e^{i2\pi\nu}\right)\right| & \leq\beta\left|R\left(e^{i2\pi\nu}\right)\right|+\left(1-\beta\right)\left|Q_{j,\star}\left(e^{i2\pi\nu}\right)\right|\\
 & <\beta\delta+\left(1-\beta\right).
\end{align*}
Consequently, $\left|Q\right|$ is smaller than $1$ on $\Gamma_{\textrm{alias}}$
as long as $\beta>0$.

It remains to prove that $\left|Q\right|$ can also be bounded by
$1$ in the rest of the torus $\Gamma_{\textrm{far}}=\overline{\mathbb{T}\backslash\left(\Gamma_{\textrm{true}}\cup\Gamma_{\textrm{alias}}\right)}$.
Let by $\mathcal{M}_{\textrm{far}}\left(R\right)$ and $\mathcal{M}_{\textrm{far}}\left(Q_{j,\star}\right)$
be the respective suprema of $R$ and $Q_{j,\star}$ over $\Gamma_{\textrm{far}}$.
$\Gamma_{\textrm{far}}$ is a closed set, and thus compact. It comes
that the suprema of $R$ and $Q$ are reached in some points inside
$\Gamma_{\textrm{far}}$. Moreover introducing the suprema of $Q_{j,\star}$
over this set
\[
\mathcal{M}_{\textrm{far}}\left(Q_{j,\star}\right)=\sup_{\nu\in\Gamma_{\textrm{far}}}\left\{ \left|Q_{j,\star}\left(e^{i2\pi\nu}\right)\right|\right\} <1,
\]
since $\tilde{\Omega}_{j}\nsubseteq\Gamma_{\textrm{far}}$, leads
to
\begin{align*}
\forall\nu\in\Gamma_{\textrm{far}},\quad\left|Q_{\star}\left(e^{i2\pi\nu}\right)\right| & \leq\beta\left|R\left(e^{i2\pi\nu}\right)\right|+\left(1-\beta\right)\left|Q_{j,\star}\left(e^{i2\pi\nu}\right)\right|\\
 & <\beta\mathcal{M}_{\textrm{far}}\left(R\right)+\left(1-\beta\right)\mathcal{M}_{\textrm{far}}\left(Q_{j,\star}\right)
\end{align*}
for all $\nu\in\Gamma_{\textrm{far}}$, and thus $\left|Q_{\star}\left(e^{i2\pi\nu}\right)\right|<1$
can be achieved everywhere on $\Gamma_{\textrm{far}}$ provided a
choice of $\beta$ verifying
\[
\beta<\frac{1-\mathcal{M}_{\textrm{far}}\left(Q_{j,\star}\right)}{\mathcal{M}_{\textrm{far}}\left(R\right)-\mathcal{M}_{\textrm{far}}\left(Q_{j,\star}\right)}.
\]

We conclude that for any coefficient $\beta$ satisfying
\[
0<\beta<\min\left\{ \frac{l_{j}\eta}{\mathcal{M}_{\baro}^{\prime\prime}\left(R\right)+l_{j}\eta},\frac{1-\mathcal{M_{\textrm{far}}}\left(Q_{j,\star}\right)}{\mathcal{M}_{\textrm{far}}\left(R\right)-\mathcal{M}_{\textrm{far}}\left(Q_{j,\star}\right)}\right\} ,
\]
the polynomial $Q_{\star}$ meet the conditions (\ref{eq:DualCertificateCondition})
and thus qualifies as a dual certificate.
\end{IEEEproof}

\section{Proof of Proposition \ref{prop:ExistenceOfCommonSupportingGrid}\label{sec:Proof-Existence of common grid}}

\subsection{Existence of a common grid}

Suppose that $\mathcal{A}_{\splus}$ is a common supporting grid for
the set of arrays $\mathbb{A}$. Relation (\ref{eq:CommonSupportingGrid-Definition})
ensures
\begin{equation}
\forall j\in\left\llbracket 1,p\right\rrbracket ,\;\forall k\in\left\llbracket 0,n_{j}-1\right\rrbracket ,\;\exists q_{j}\left[k\right]\in\left\llbracket 0,n_{\splus}-1\right\rrbracket \quad\text{s.t.}\quad\frac{1}{f_{j}}\left(k-\gamma_{j}\right)=\frac{1}{f_{\splus}}\left(q_{j}\left[k\right]-\gamma_{\splus}\right),\label{eq:CommonGridExistence}
\end{equation}
whereby each integer $q_{j}\left[k\right]$ represents the position
of the $k^{\textrm{th}}$ samples of the $j^{\textrm{th}}$ grid in
the common grid. By subtracting two instances of (\ref{eq:CommonGridExistence})
applied to the grid $j$ and for the samples of order $k$ and $k+1$
one gets
\[
\forall j\in\left\llbracket 1,p\right\rrbracket ,\forall k\in\left\llbracket 0,n_{j}-1\right\rrbracket ,\quad\frac{f_{\splus}}{f_{j}}=q_{j}\left[k+1\right]-q_{j}\left[k\right]\triangleq l_{j},
\]
where $\left\{ l_{j}\right\} _{j\in\left\llbracket 1,p\right\rrbracket }$
are positive integers since $q_{j}$ is an increasing sequence for
all $j\in\left\llbracket 1,p\right\rrbracket $. It comes that $\left\{ q_{j}\right\} _{j\in\left\llbracket 1,p\right\rrbracket }$
are $p$ arithmetic progressions with respective increment $l_{j}$
\[
\forall j\in\left\llbracket 1,p\right\rrbracket ,\forall k\in\left\llbracket 0,n_{j}-1\right\rrbracket ,\quad q_{j}\left[k\right]=q_{j}\left[0\right]+l_{j}k.
\]
Reporting those results in Equation (\ref{eq:CommonGridExistence})
leads to
\[
\forall j\in\left\llbracket 1,p\right\rrbracket ,\quad\gamma_{\splus}=q_{j}\left[0\right]+l_{j}\gamma_{j}.
\]
Letting $a_{j}=-q_{j}\left[0\right]$ for all $j\in\left\llbracket 1,p\right\rrbracket $
proofs the necessity part.

On the other hand, suppose now the existence of positive integers
$\left\{ l_{j}\right\} \mathbb{\in\mathbb{N}}^{p}$ and integers $\left\{ a_{j}\right\} \mathbb{\in\mathbb{Z}}^{p}$
such that the relations
\begin{equation}
\begin{cases}
f_{\splus}=l_{j}f_{j}, & \forall j\in\left\llbracket 1,p\right\rrbracket \\
\gamma_{\splus}=l_{j}\gamma_{j}-a_{j}, & \forall j\in\left\llbracket 1,p\right\rrbracket ,
\end{cases}\label{eq:GridExpansion}
\end{equation}
hold for some $f_{\splus}\in\mathbb{R}^{+}$ and $\gamma_{\splus}\in\mathbb{R}$.
It comes
\begin{align}
\forall j\in\left\llbracket 1,p\right\rrbracket ,\forall k\in\left\llbracket 0,n_{j}-1\right\rrbracket ,\quad\frac{1}{f_{j}}\left(k-\gamma_{j}\right) & =\frac{1}{f_{j}}\left(k-l_{j}a_{j}-l_{j}\gamma_{\splus}\right)\nonumber \\
 & =\frac{1}{f_{\splus}}\left(l_{j}k-a_{j}-\gamma_{\splus}\right).\label{eq:CommonGrid-Progression}
\end{align}
Defining the quantities
\begin{equation}
\begin{cases}
q_{j}\left[k\right]=l_{j}k-a_{j}, & \forall j\in\left\llbracket 1,p\right\rrbracket \\
n_{\splus}\geq\max_{j\in\left\llbracket 1,p\right\rrbracket }\left\{ q_{j}\left[n_{j}-1\right]\right\} ,
\end{cases}\label{eq:CommonGrid-Parametrization}
\end{equation}
ensures that the grid $\mathcal{A}_{\splus}=\left(f_{\splus},\gamma_{\splus},n_{\splus}\right)$
supports the system defined by $\mathbb{A}$. This achieves the sufficiency
part, and thus the characterization of the existence of a common grid.

\subsection{Conditions for minimality}

Suppose that $\mathbb{A}$ admits a common grid, it is clear that
exactly one element of $\mathcal{C}\left(\mathbb{A}\right)$ reaches
the minimal order $n_{\baro}$. Denote by $\mathcal{A}_{\baro}=\left(f_{\baro},\gamma_{\baro},n_{\baro}\right)$
this element. Moreover, denote by $\left\{ l_{j}\right\} \mathbb{\in\mathbb{N}}^{p}$
and $\left\{ a_{j}\right\} \mathbb{\in\mathbb{Z}}^{p}$ the elements
characterizing the grid expansion of $\mathbb{A}$ onto $\mathcal{A}_{\baro}$
defined in (\ref{eq:GridExpansion}), and let $\delta=\gcd\left(\left\{ a_{j}\right\} _{j\in\left\llbracket 1,p\right\rrbracket }\cup\left\{ l_{j}\right\} _{j\in\left\llbracket 1,p\right\rrbracket }\right)$.
By (\ref{eq:CommonGrid-Progression}), one has
\[
\forall j\in\left\llbracket 1,p\right\rrbracket ,\forall k\in\left\llbracket 0,n_{j}-1\right\rrbracket ,\quad\frac{1}{f_{j}}\left(k-\gamma_{j}\right)=\frac{\delta}{f_{\baro}}\left(\frac{l_{j}}{\delta}k-\frac{a_{j}}{\delta}-\frac{\gamma_{\baro}}{\delta}\right),
\]
Thus the grid $\mathcal{A}_{\baro}=\left(\frac{f_{\baro}}{\delta},\frac{\gamma_{\baro}}{\delta},\left\lceil \frac{n_{\baro}}{\delta}\right\rceil \right)$
supports $\mathbb{A}$ and belongs to $\mathcal{C}\left(\mathcal{\mathbb{A}}\right)$.
My minimality of $\mathcal{A}_{\baro}$ one has $\left\lceil \frac{n_{\baro}}{\delta}\right\rceil \geq n_{\baro}$
and we conclude that $\delta=1$. Moreover, the minimality implies
that the first and the last samples of the grid $\mathcal{A}_{\baro}$
must be acquired by an element of $\mathbb{A}$, otherwise the shorter
grids $\mathcal{A}_{\baro}=\left(f_{\baro},\gamma_{\baro}-1,n_{\baro}-1\right)$,
or $\mathcal{A}_{\baro}=\left(f_{\baro},\gamma_{\baro}+1,n_{\baro}-1\right)$
would also support $\mathbb{A}$. Using (\ref{eq:CommonGrid-Parametrization})
\[
\begin{cases}
\forall j\in\left\llbracket 1,p\right\rrbracket , & \gamma_{\baro}=l_{j}\gamma_{j}-a_{j}\\
\exists j\in\left\llbracket 1,p\right\rrbracket , & a_{j}=0\\
\forall j\in\left\llbracket 1,p\right\rrbracket , & a_{j}\leq0,
\end{cases}
\]
which implies $\gamma_{\baro}=\max_{j\in\left\llbracket 1,p\right\rrbracket }\left\{ l_{j}\gamma_{j}\right\} $,
ensuring that the conditions describing the minimal grid stated in
Proposition \ref{prop:ExistenceOfCommonSupportingGrid} are necessary.

\sloppy For the sufficiency, consider the grid $\mathcal{A}_{\baro}=\left(f_{\baro},\gamma_{\baro},n_{\baro}\right)$
of $\mathcal{C}\left(\mathbb{A}\right)$ where $\gamma_{\baro}=\max_{j\in\left\llbracket 1,p\right\rrbracket }\left\{ l_{j}\gamma_{j}\right\} $
and with expansion parameters $\left\{ l_{j}\right\} \mathbb{\in\mathbb{N}}^{p}$
and $\left\{ a_{j}\right\} \mathbb{\in\mathbb{Z}}^{p}$ satisfying
$\gcd\left(\left\{ a_{j}\right\} \cup\left\{ l_{j}\right\} ,\;j\in\left\llbracket 1,p\right\rrbracket \right)=1$.
Let $\mathcal{A^{\prime}}=\left(f^{\prime},\gamma^{\prime},n^{\prime}\right)\in\mathcal{C}\left(\mathbb{A}\right)$
be any other grid and let by $\delta^{\prime}$ its corresponding
greatest common divisor. $\delta^{\prime}$ devises every integer
linear combination of $\left\{ a_{j}\right\} \cup\left\{ l_{j}\right\} $,
and in particular every elements of the set $\left\{ l_{j}k_{j}-a_{j}:\;j\in\left\llbracket 1,p\right\rrbracket ,k_{j}\in\left\llbracket 0,n_{j}-1\right\rrbracket \right\} $.
Therefore $\left(f^{\prime},\gamma^{\prime}\right)$ is identifiable
to $\left(\delta^{\prime}f_{\baro},\delta^{\prime}\gamma_{\baro}-b\right)$
for some $b\in\mathbb{Z}$. Moreover since $\gamma_{\baro}$ is maximum,
the grid $\mathcal{A}_{\baro}$ samples an element of $\mathbb{A}$
at index $0$, and thus $\mathcal{A}^{\prime}\in\mathcal{C}\left(\mathbb{A}\right)$
if only and only if $b\geq0$. Finally it comes from (\ref{eq:CommonGrid-Parametrization})
that $n^{\prime}$ must satisfy
\begin{align*}
n^{\prime} & \geq\max_{j\in\left\llbracket 1,p\right\rrbracket }\left\{ q_{j}^{\prime}\left[n_{j}-1\right]\right\} \\
 & \geq\max_{j\in\left\llbracket 1,p\right\rrbracket }\left\{ \delta^{\prime}l_{j}\left(n_{j}-1\right)-\delta^{\prime}l_{j}a_{j}+b\right\} \\
 & \geq\max_{j\in\left\llbracket 1,p\right\rrbracket }\left\{ l_{j}\left(n_{j}-1\right)-a_{j}\right\} \\
 & \geq n_{\baro},
\end{align*}
\fussy demonstrating the sufficiency part, and concluding the proof
of Proposition \ref{prop:ExistenceOfCommonSupportingGrid}.

\bibliographystyle{IEEEtran}
\bibliography{BibTeX}

\begin{IEEEbiographynophoto}{Maxime Ferreira Da Costa}
\end{IEEEbiographynophoto}

\begin{IEEEbiographynophoto}{Wei Dai}
\end{IEEEbiographynophoto}

\end{document}